\documentclass{article}

\PassOptionsToPackage{numbers, compress}{natbib}

 \usepackage[preprint]{neurips_2026}

\usepackage[utf8]{inputenc} 
\usepackage[T1]{fontenc}    
\usepackage{hyperref}       
\usepackage{url}            
\usepackage{booktabs}       
\usepackage{amsfonts}       
\usepackage{nicefrac}       
\usepackage{microtype}      
\usepackage{xcolor}         
\usepackage{colortbl}       

\newcommand{\heatcell}[1]{%
  \ifdim #1pt > 79.5pt \cellcolor{green!40}#1%
  \else\ifdim #1pt > 59.5pt \cellcolor{green!20}#1%
  \else\ifdim #1pt > 39.5pt \cellcolor{yellow!30}#1%
  \else\ifdim #1pt > 19.5pt \cellcolor{orange!30}#1%
  \else \cellcolor{red!30}#1%
  \fi\fi\fi\fi
}
\newcommand{\heatcellpm}[2]{%
  \ifdim #1pt > 79.5pt \cellcolor{green!40}#1{\tiny$\pm$#2}%
  \else\ifdim #1pt > 59.5pt \cellcolor{green!20}#1{\tiny$\pm$#2}%
  \else\ifdim #1pt > 39.5pt \cellcolor{yellow!30}#1{\tiny$\pm$#2}%
  \else\ifdim #1pt > 19.5pt \cellcolor{orange!30}#1{\tiny$\pm$#2}%
  \else \cellcolor{red!30}#1{\tiny$\pm$#2}%
  \fi\fi\fi\fi
}
\usepackage[disable]{todonotes}
\usepackage{booktabs}
\usepackage{float}
\usepackage{afterpage}
\usepackage{subcaption}
\usepackage{listings}

\definecolor{codegreen}{rgb}{0.25,0.49,0.25}
\definecolor{codegray}{rgb}{0.5,0.5,0.5}
\definecolor{codepurple}{rgb}{0.58,0,0.82}
\definecolor{codeblue}{rgb}{0.13,0.13,0.7}
\definecolor{codebg}{rgb}{0.96,0.96,0.96}

\usepackage{tikz}
\usetikzlibrary{arrows.meta, positioning, shapes.geometric}
\usepackage{pgfplots}
\pgfplotsset{compat=1.16}

\title{Teaching LLMs Program Semantics\\via Symbolic Execution Traces}

\author{%
  Jonas Bayer\thanks{Work done during an internship at Amazon Web Services.} \\
  University of Cambridge\\
  \And
  Stefan Zetzsche\thanks{Corresponding author: \texttt{stefanze@amazon.co.uk}} \\
  Amazon Web Services\\
  \And
  Olivier Bouissou\\
  Amazon Web Services\\
  \And
  Remi Delmas\\
  Amazon Web Services\\
  \And
  Michael Tautschnig\\
  Amazon Web Services\\
  \And
  Soonho Kong\\
  Amazon Web Services\\
}

\begin{document}


\maketitle


\begin{abstract}
We introduce an evaluation framework of 500 C verification tasks across five property types (memory safety, overflow, termination, reachability, data races) built on SV-COMP 2025, and evaluate 14 models across six families. We find that high overall accuracy masks a critical weakness: while most models reliably confirm properties hold, violation detection varies widely and degrades sharply with program length. To close this gap, we train on formal verification artifacts: running the Soteria symbolic execution engine on generic open-source C code and using the resulting traces for continued pretraining of Qwen3-8B. Just ${\sim}$3,000 bug traces combined with chain-of-thought reasoning at inference time improve violation detection by over 17 percentage points, producing one of the most balanced accuracy profiles among evaluated models. On violation detection, the trained 8B model outperforms the 4$\times$ larger Qwen3-32B without thinking and approaches it in overall accuracy. The interaction between trace training and chain-of-thought is superadditive: neither alone provides meaningful gains, but their combination does. Improvements transfer across all five property types, including ones the training traces do not target. Our 28 configurations confirm the gains stem from trace semantics, not code volume, and that trace curation and format matter.
\end{abstract}


\section{Introduction}

How well do LLMs understand low-level code? For higher-level languages, code understanding evaluations are approaching saturation \citep{xie2025core}, but C requires reasoning about memory safety, pointer arithmetic, and undefined behavior, where models remain less reliable \citep{qiu2025locobench}. Yet there is no established evaluation framework that isolates this kind of semantic reasoning for C.

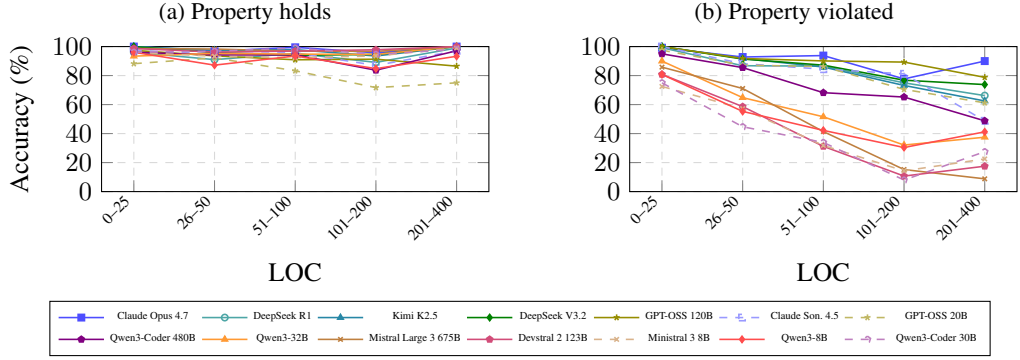
\begin{figure}[t!]
\centering
\begin{subfigure}{0.48\columnwidth}
\centering
\subcaption{Property holds}
\begin{tikzpicture}
\begin{axis}[
    width=\textwidth,
    height=3.5cm,
    ylabel={Accuracy (\%)},
    xlabel={LOC},
    ymin=0, ymax=100,
    symbolic x coords={0--25,26--50,51--100,101--200,201--400},
    xtick=data,
    xticklabel style={font=\tiny, rotate=45, anchor=east},
    grid=major,
    grid style={dashed, gray!30},
    every axis plot/.append style={semithick, mark size=1.2pt},
    legend to name=sharedlegend,
    legend style={font=\fontsize{4}{5}\selectfont, legend columns=7},
]
\addplot[color=blue!70, mark=square*] coordinates {(0--25,100.0) (26--50,96.8) (51--100,99.5) (101--200,95.7) (201--400,100.0)};
\addplot[color=teal!70, mark=o] coordinates {(0--25,96.1) (26--50,91.0) (51--100,95.3) (101--200,89.1) (201--400,99.0)};
\addplot[color=cyan!60!black, mark=triangle*] coordinates {(0--25,100.0) (26--50,97.4) (51--100,97.9) (101--200,93.5) (201--400,100.0)};
\addplot[color=green!50!black, mark=diamond*] coordinates {(0--25,100.0) (26--50,95.9) (51--100,94.0) (101--200,94.6) (201--400,100.0)};
\addplot[color=olive, mark=star] coordinates {(0--25,97.4) (26--50,93.6) (51--100,90.9) (101--200,91.3) (201--400,86.5)};
\addplot[color=blue!40, mark=square, dashed] coordinates {(0--25,97.4) (26--50,94.8) (51--100,95.8) (101--200,87.0) (201--400,97.1)};
\addplot[color=olive!60, mark=star, dashed] coordinates {(0--25,88.2) (26--50,92.2) (51--100,83.3) (101--200,71.7) (201--400,75.0)};
\addplot[color=violet, mark=pentagon*] coordinates {(0--25,96.1) (26--50,94.2) (51--100,94.3) (101--200,83.7) (201--400,97.1)};
\addplot[color=orange!80, mark=triangle*] coordinates {(0--25,93.4) (26--50,95.3) (51--100,94.8) (101--200,94.6) (201--400,100.0)};
\addplot[color=brown, mark=x] coordinates {(0--25,98.7) (26--50,98.3) (51--100,96.9) (101--200,96.7) (201--400,100.0)};
\addplot[color=purple!70, mark=pentagon*] coordinates {(0--25,98.7) (26--50,96.2) (51--100,96.9) (101--200,97.8) (201--400,100.0)};
\addplot[color=brown!60, mark=x, dashed] coordinates {(0--25,96.1) (26--50,93.3) (51--100,92.4) (101--200,93.5) (201--400,99.0)};
\addplot[color=red!70, mark=diamond*] coordinates {(0--25,96.1) (26--50,87.2) (51--100,93.5) (101--200,84.8) (201--400,93.3)};
\addplot[color=violet!50, mark=pentagon, dashed] coordinates {(0--25,96.1) (26--50,96.8) (51--100,99.0) (101--200,94.6) (201--400,98.1)};
\legend{Claude Opus 4.7, DeepSeek R1, Kimi K2.5, DeepSeek V3.2, GPT-OSS 120B, Claude Son.\ 4.5, GPT-OSS 20B, Qwen3-Coder 480B, Qwen3-32B, Mistral Large 3 675B, Devstral 2 123B, Ministral 3 8B, Qwen3-8B, Qwen3-Coder 30B}
\end{axis}
\end{tikzpicture}
\end{subfigure}
\hfill
\begin{subfigure}{0.48\columnwidth}
\centering
\subcaption{Property violated}
\begin{tikzpicture}
\begin{axis}[
    width=\textwidth,
    height=3.5cm,
    xlabel={LOC},
    ymin=0, ymax=100,
    symbolic x coords={0--25,26--50,51--100,101--200,201--400},
    xtick=data,
    xticklabel style={font=\tiny, rotate=45, anchor=east},
    grid=major,
    grid style={dashed, gray!30},
    every axis plot/.append style={semithick, mark size=1.2pt},
]
\addplot[color=blue!70, mark=square*] coordinates {(0--25,99.2) (26--50,92.8) (51--100,93.8) (101--200,77.7) (201--400,90.0)};
\addplot[color=teal!70, mark=o] coordinates {(0--25,99.2) (26--50,86.8) (51--100,86.5) (101--200,75.0) (201--400,66.2)};
\addplot[color=cyan!60!black, mark=triangle*] coordinates {(0--25,100.0) (26--50,91.8) (51--100,85.9) (101--200,73.2) (201--400,62.5)};
\addplot[color=green!50!black, mark=diamond*] coordinates {(0--25,100.0) (26--50,91.4) (51--100,87.2) (101--200,76.8) (201--400,73.8)};
\addplot[color=olive, mark=star] coordinates {(0--25,100.0) (26--50,91.8) (51--100,90.1) (101--200,89.3) (201--400,78.8)};
\addplot[color=blue!40, mark=square, dashed] coordinates {(0--25,100.0) (26--50,88.2) (51--100,84.4) (101--200,81.2) (201--400,48.8)};
\addplot[color=olive!60, mark=star, dashed] coordinates {(0--25,97.5) (26--50,87.2) (51--100,86.2) (101--200,70.5) (201--400,61.2)};
\addplot[color=violet, mark=pentagon*] coordinates {(0--25,95.0) (26--50,85.5) (51--100,68.2) (101--200,65.2) (201--400,48.8)};
\addplot[color=orange!80, mark=triangle*] coordinates {(0--25,90.0) (26--50,64.8) (51--100,51.6) (101--200,32.1) (201--400,37.5)};
\addplot[color=brown, mark=x] coordinates {(0--25,85.8) (26--50,71.1) (51--100,41.4) (101--200,15.2) (201--400,8.8)};
\addplot[color=purple!70, mark=pentagon*] coordinates {(0--25,80.8) (26--50,58.6) (51--100,31.2) (101--200,10.7) (201--400,17.5)};
\addplot[color=brown!60, mark=x, dashed] coordinates {(0--25,72.5) (26--50,57.2) (51--100,31.8) (101--200,14.3) (201--400,22.5)};
\addplot[color=red!70, mark=diamond*] coordinates {(0--25,80.8) (26--50,55.3) (51--100,42.2) (101--200,30.4) (201--400,41.2)};
\addplot[color=violet!50, mark=pentagon, dashed] coordinates {(0--25,75.0) (26--50,44.7) (51--100,33.9) (101--200,8.0) (201--400,27.5)};
\end{axis}
\end{tikzpicture}
\end{subfigure}

\vspace{0.1cm}
\centering\ref{sharedlegend}

\caption{Baseline accuracy on our benchmark by program length for all models. Holds accuracy (a) remains stable for most models. Violation detection (b) degrades sharply with program length.}
\label{fig:length-decay-main}
\end{figure}

We introduce an evaluation framework built on SV-COMP 2025 \citep{beyer2025improvements}, the Software Verification Competition. Given a C program and a property (memory safety, absence of overflows, termination, reachability, or absence of data races), models must determine whether the property holds or is violated, with ground truth established by formal verification tools. We evaluate 14 models across six families and find that while most models achieve over 90\% accuracy on confirming properties hold, violation detection is far weaker: four of fourteen models correctly identify fewer than 50\% of actual violations, and accuracy drops sharply with program length, with one model falling below 10\% on programs of just 100--200 lines (Figure~\ref{fig:length-decay-main}). Notably, this does not correlate straightforwardly with model size or code specialization; for instance, GPT-OSS-20B detects violations more reliably than Mistral Large 3 675B and Qwen3-Coder-480B, suggesting that training data and objectives play a larger role than scale. A failure analysis (Appendix~\ref{app:failure-example}) reveals that models fail by missing triggering edge cases, making incorrect logical or arithmetic reasoning steps, or identifying the violation but incorrectly concluding it is benign.

To improve violation detection, we explore using formal verification artifacts as training data. Prior work combining LLMs with formal methods has largely focused on inference time; the training-time direction is less explored. Existing efforts have targeted theorem proving with Lean proofs \citep{moura2021lean, ren2025deepseek} and training on concrete Python execution traces \citep{copet2025cwm}. We extend this idea to C, where a mature ecosystem of symbolic execution tools exists, yet their artifacts remain untapped for LLM training.

We focus on symbolic execution, which reasons about all possible inputs simultaneously rather than sampling concrete values. Symbolic execution artifacts capture program semantics in a form well-suited for training: they trace execution paths, track memory states, and reveal the precise conditions under which properties are violated. We use Soteria \citep{ayoun2025soteria}, an open-source symbolic execution engine that is fast, produces detailed traces, and supports both C and Rust. Starting from the GitHub CodeParrot dataset \citep{codeparrot}, we filter to 1 million self-contained C files and run Soteria on each, yielding approximately 34,000 usable traces. The reduction stems from timeout and parser limitations in Soteria; we expect the yield to improve as the tooling matures.

We conduct extensive ablation studies on Qwen3-8B to understand how different aspects of the training data affect performance. Our main finding is that filtering to manifest bug traces (where Soteria proves a violation occurs on all feasible inputs), combined with chain-of-thought reasoning at inference time, yields a 17.9 percentage point improvement on detecting property violations. The reported holds accuracy drops from 90.7\% to 78.0\%, but this is driven entirely by responses that exhaust the output token budget; on parseable responses, holds accuracy is preserved (90.8\% vs.\ 91.2\% baseline) while violated accuracy improves by +26 pp. This gain depends on careful curation: training on all Soteria traces degrades performance, and training on the underlying C files alone also hurts, confirming that improvement derives from the semantic information in Soteria's traces rather than simply increased code exposure. Importantly, the training data is entirely independent of our benchmark, drawn from arbitrary open-source C code, demonstrating that improvements transfer to unseen verification tasks.

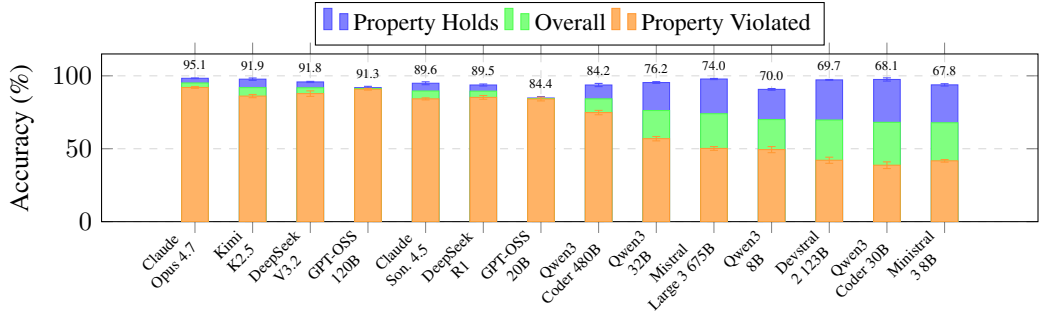
\begin{figure}[t]
\centering
\begin{tikzpicture}
\begin{axis}[
    ybar,
    width=\columnwidth,
    height=3.8cm,
    ylabel={Accuracy (\%)},
    ymin=0, ymax=115,
    xtick={1,2,3,4,5,6,7,8,9,10,11,12,13,14},
    xticklabels={
        {Claude\\Opus 4.7},
        {Kimi\\K2.5},
        {DeepSeek\\V3.2},
        {GPT-OSS\\120B},
        {Claude\\Son.\ 4.5},
        {DeepSeek\\R1},
        {GPT-OSS\\20B},
        {Qwen3\\Coder 480B},
        {Qwen3\\32B},
        {Mistral\\Large 3 675B},
        {Qwen3\\8B},
        {Devstral\\2 123B},
        {Qwen3\\Coder 30B},
        {Ministral\\3 8B}
    },
    xticklabel style={font=\tiny, align=center, rotate=45, anchor=east},
    legend style={at={(0.5,1.05)}, anchor=south, font=\small, legend columns=3},
    enlarge x limits=0.08,
    grid=major,
    grid style={dashed, gray!30},
    xmin=0.5, xmax=14.5,
]
\addplot[bar width=10pt, bar shift=0, fill=blue!50, draw=blue!70,
    nodes near coords,
    nodes near coords style={font=\fontsize{5}{6}\selectfont, anchor=south},
    point meta=explicit symbolic,
    error bars/.cd, y dir=both, y explicit, error bar style={line width=0.4pt, color=blue!70}, error mark options={rotate=90, mark size=1.5pt, color=blue!70}] coordinates {(1,98.3) +- (0,0.2) [95.1] (2,97.7) +- (0,0.8) [91.9] (3,95.8) +- (0,0.4) [91.8] (4,91.9) +- (0,0.9) [91.3] (5,94.9) +- (0,1.0) [89.6] (6,93.7) +- (0,0.7) [89.5] (7,84.8) +- (0,1.0) [84.4] (8,93.7) +- (0,1.1) [84.2] (9,95.4) +- (0,0.6) [76.2] (10,97.8) +- (0,0.4) [74.0] (11,90.7) +- (0,0.6) [70.0] (12,97.2) +- (0,0.3) [69.7] (13,97.5) +- (0,1.1) [68.1] (14,93.8) +- (0,0.8) [67.8]
};
\addplot[bar width=10pt, bar shift=0, fill=green!50, draw=green!70] coordinates {(1,95.1) (2,91.9) (3,91.8) (4,91.3) (5,89.6) (6,89.5) (7,84.4) (8,84.2) (9,76.2) (10,74.0) (11,70.0) (12,69.7) (13,68.1) (14,67.8)
};
\addplot[bar width=10pt, bar shift=0, fill=orange!60, draw=orange!80,
    error bars/.cd, y dir=both, y explicit, error bar style={line width=0.4pt, color=orange!80}, error mark options={rotate=90, mark size=1.5pt, color=orange!80}] coordinates {(1,92.0) +- (0,0.6) (2,86.1) +- (0,1.1) (3,87.8) +- (0,1.9) (4,90.8) +- (0,0.6) (5,84.2) +- (0,0.7) (6,85.2) +- (0,1.3) (7,84.1) +- (0,1.3) (8,74.8) +- (0,1.5) (9,56.9) +- (0,1.5) (10,50.2) +- (0,1.3) (11,49.4) +- (0,2.1) (12,42.1) +- (0,2.1) (13,38.7) +- (0,2.3) (14,41.7) +- (0,0.9)
};
\legend{Property Holds, Overall, Property Violated}
\end{axis}
\end{tikzpicture}
\caption{Baseline accuracy on our benchmark, sorted by overall performance. Numbers above each bar show overall accuracy (\%). All models detect property holds more reliably than violations, with the gap widening dramatically for the Mistral and Qwen families regardless of scale. Figure~\ref{fig:length-decay-main} shows that violation detection degrades sharply already within the 100--200 LOC range.}
\label{fig:baseline-bar}
\end{figure}

Our contributions are:
\begin{itemize}
\item An evaluation framework of 500 verification tasks across five property types, with ground truth from SV-COMP 2025. A baseline evaluation of 14 models reveals that violation detection varies widely --- from 92\% for the best model down to below 40\% --- a deficit often masked by strong performance on holds cases, which inflates overall accuracy.
\item A scalable, fully automated, annotation-free training pipeline: run Soteria on generic open-source C code, filter to manifest bug traces, and continue pretraining. Just 3,208 traces and {$\sim$}10 minutes of training improve Qwen3-8B violation detection by +17.9 pp while also improving overall accuracy. Although the training traces primarily target memory safety and overflows, violation detection improves across all five evaluation properties, suggesting the model acquires general verification reasoning.
\item Extensive ablations (28 configurations) showing that gains are driven by the semantic content of symbolic execution traces, not code volume, and that trace format, curation, and thinking mode interact non-obviously: CPT (continued pretraining) and thinking are individually weak but superadditive in combination (+17.9 pp vs.\ +7.3 pp and $-$1.4 pp alone), and preserving the original structured trace format is critical for chain-of-thought to succeed.
\end{itemize}


\section{Related Work}

\paragraph{LLMs for Code Understanding.}
Benchmarks for semantic code understanding remain limited. CRUXEval \citep{gu2024cruxeval} evaluates input-output prediction for Python but not low-level languages; CoRe \citep{xie2025core} tests data and control dependencies across Python, C, and Java but is approaching saturation; LoCoBench \citep{qiu2025locobench} evaluates long-context understanding across 10 languages with lower C/C++ performance. In security, recent work observes that models struggle more with detecting vulnerabilities than classifying benign code \citep{dubniczky2025castle, li2025everything}, consistent with the holds--violated asymmetry we document. Concurrent with our work, Sultan et al.\ \citep{sultan2026llms} evaluate LLMs on SV-COMP 2025 termination tasks, and Wei et al.\ \citep{wei2025invbench} benchmark invariant synthesis.

\paragraph{Training on Program Behavior Artifacts.}
Training on Lean proofs has improved mathematical reasoning \citep{ren2025deepseek,achim2025aristotle,chen2025seed,wang2025kimina,hubert2025olympiad}. Verified code generation is active, with training pipelines \citep{chen2024automated, baksys2025atlas} and benchmarks \citep{ye2025verina, thakur2025clever, bursuc2025benchmark} for Lean, Dafny, and Verus. Sakharova et al.\ \citep{sakharova2025integrating} use CrossHair (Python symbolic execution) to improve RL reward signals. Several approaches train on concrete execution traces: ET-CoT \citep{eguchi2025making} fine-tunes on traces to predict outcomes, NExT \citep{ni2024next} uses traces for repair, and Chen et al.\ \citep{chen2025chain} show execution supervision promotes general reasoning. CWM \citep{copet2025cwm} and others \citep{liu2023code, armengol2025cannot} train on Python execution traces. These focus on high-level languages and concrete execution; we train on symbolic execution traces for C, reasoning over all possible inputs.

\paragraph{Symbolic Execution and Program Verification.}
SV-COMP~\cite{beyer2025improvements} is an annual competition benchmarking verification tools on properties like memory safety, overflow absence, termination, and reachability. For C, prominent tools include CBMC \citep{kroening2014cbmc}, Infer \citep{calcagno2015moving}, and UAutomizer \citep{heizmann2018ultimate}, the winner in many categories of SV-COMP 2025; Kani \citep{kani} targets Rust, and Soteria \citep{ayoun2025soteria}, while language-agnostic, currently offers backends for C and Rust. We use SV-COMP for evaluation and Soteria for training data due to its speed and detailed logs.

\paragraph{Hybrid Approaches.}
Tool-augmented approaches combine LLMs with formal methods at inference time. Lemur \citep{wu2024lemur} integrates LLMs with automated reasoners for program verification, using LLMs to generate invariants and annotations. SatLM \citep{ye2023satlm} has LLMs generate declarative specifications solved by automated theorem provers. SkipAnalyzer \citep{mohajer2023skipanalyzer} combines LLMs with Infer for bug detection and false-positive filtering. M{\"a}chtle et al.\ \citep{machtle2025trace} use execution traces to minimize code context for vulnerability prediction. Fedoseev et al.\ \citep{fedoseev24constraint} train models to generate Z3 specifications for constraint solving. Maaz et al.\ \citep{maaz2025agentic} use an LLM agent to synthesize and execute property-based tests across the Python ecosystem. These approaches require formal tools at inference time; we use formal tool artifacts purely for training, requiring no tool invocation at inference.


\section{Evaluation Framework}

To measure how well LLMs understand low-level code semantics, we need a benchmark that isolates semantic reasoning from other capabilities like code generation. We construct an evaluation framework where models perform binary classification: given a C program and a property, determine whether the property holds or is violated.

\subsection{Benchmark Construction}
\label{sec:benchmark}

\paragraph{Source Data.}
We use SV-COMP 2025 \citep{beyer2025improvements}, which contains approximately 40,000 verification tasks across a repository of roughly 76,000 C files, with ground truth established by formal verification tools.\footnote{SV-COMP 2025 benchmarks are publicly available on GitLab and Zenodo. The competition infrastructure is licensed under Apache 2.0.} An earlier evaluation on CBMC regression tests \citep{cbmc-regression} showed models performed too well to differentiate improvements (Appendix~\ref{app:cbmc}).

\begin{figure}[t]
\centering
\begin{lstlisting}[language=C, basicstyle=\fontsize{7}{8.5}\selectfont\ttfamily]
typedef enum {false, true} bool;
extern unsigned int __VERIFIER_nondet_uint(void);

int main()
{
    unsigned int x;
    x = __VERIFIER_nondet_uint();
    if (x > 0) {
        while (x != 0) {
            x = x - 2;
        }
    }
    return 0;
}
\end{lstlisting}
\caption{Example from our benchmark: does this program always terminate? When \texttt{x} is odd, unsigned wrapping causes an infinite cycle. Mistral Large 3 675B identifies the wrapping but incorrectly concludes the sequence must eventually reach zero (Appendix~\ref{app:failure-example}).}
\label{fig:example-program}
\end{figure}

\paragraph{Preprocessing.}
SV-COMP 2025 was designed for evaluating verification tools, not LLMs, so some preprocessing is needed. We restrict to single-file tasks that do not require custom header files (standard library includes like \texttt{pthread.h} and \texttt{stdlib.h} are permitted), reducing the pool to about 15,000 files. This constraint could be relaxed in future work by providing multi-file context to the model. We apply additional filtering to ensure data quality, and strip comments from source code, as formal tools ignore them but LLMs may extract ground truth from them. The full list of preprocessing steps is provided in Appendix~\ref{app:svcomp-preprocessing}.

\paragraph{Holdout Curation.}
The original dataset contains approximately 80\% tasks where the property holds and 20\% where it is violated, so a model could achieve 80\% accuracy by always predicting ``property holds.'' SV-COMP addresses this with an asymmetric scoring scheme that penalizes incorrect claims more heavily; we instead use simple accuracy on a balanced holdout set of 500 tasks (reporting the macro-average of holds and violated accuracy, which weights each side equally regardless of the natural class distribution):
\begin{itemize}
\item 100 tasks for each of five core property categories: memory safety, absence of overflows, termination, reachability, and absence of data races
\item 50/50 split between property holds and property violated for each property
\item Varying program lengths (0--400 lines), balanced within each length bin
\end{itemize}

\paragraph{Prompt Design.}
For each property, we construct a prompt containing the C code, a property-specific description, and output format instructions, including explanations of SV-COMP primitives such as \texttt{\_\_VERIFIER\_nondet\_X()}. Full details are in Appendix~\ref{app:svcomp-prompt}; Figure~\ref{fig:example-program} shows an example. We use a relaxed answer parser that tolerates common formatting variations, avoiding the need for instruction tuning solely to match a rigid output format.

\paragraph{Scope and Limitations.}
In the standard SV-COMP setting, tools must also produce a machine-checkable witness justifying the claim. We evaluate only the binary verdict; requiring witness generation would be substantially harder \citep{sultan2026llms}.

\subsection{Baseline Performance}

We evaluate 14 models on our benchmark. Twelve are accessed via Amazon Bedrock; Qwen3-8B and Qwen3-32B are evaluated self-hosted via vLLM. All baseline evaluations use greedy decoding (temperature=0, except Claude Opus 4.7 which does not support temperature parameters) without chain-of-thought reasoning; we explore the effect of enabling thinking in Section~\ref{sec:results}. All 14 models are evaluated 4 times and we report the mean across runs; standard deviations on overall accuracy are below 1.5 percentage points for all models, confirming high reproducibility across inference runs. Variance over the broader distribution of C programs is addressed by the diversity of SV-COMP tasks, which are drawn from real software projects and designed to challenge state-of-the-art verifiers. Per-property variance can be higher due to the smaller sample size (50 tasks per property per side), so per-property comparisons should be interpreted with caution. Figure~\ref{fig:baseline-bar} visualizes the results; per-property breakdowns are in Appendix~\ref{app:svllm-baseline-detailed}. Tasks that exceed the input context window, produce unparseable responses (output budget exhaustion), or time out are all counted as incorrect.

The results reveal a consistent pattern: confirming that a property holds is easier than detecting violations. Frontier models handle both well (Claude Opus 4.7: 98.3\% holds, 92.0\% violated), but four of fourteen models correctly identify fewer than 50\% of actual violations despite all achieving over 84\% on holds. This does not correlate with model size or code specialization: GPT-OSS-20B (84.1\% violated) outperforms Qwen3-Coder-480B (74.8\%), and the entire Mistral family achieves 41--51\% on violations regardless of scale (8B to 675B). Comparing DeepSeek R1 with V3.2 shows no meaningful improvement from built-in chain-of-thought, suggesting that training data matters more than scale for verification tasks.

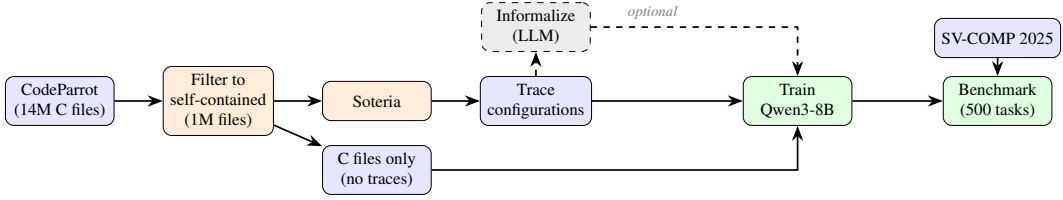
\begin{figure}[t!]
\centering
\resizebox{\columnwidth}{!}{%
\begin{tikzpicture}[
    node distance=0.6cm and 0.8cm,
    box/.style={rectangle, draw, rounded corners, minimum height=0.65cm, minimum width=1.8cm, align=center, font=\footnotesize},
    data/.style={box, fill=blue!10},
    process/.style={box, fill=orange!15},
    result/.style={box, fill=green!12},
    optional/.style={box, fill=gray!15, dashed},
    arr/.style={-{Stealth[length=2.5mm]}, thick},
    optarr/.style={-{Stealth[length=2.5mm]}, thick, dashed},
]
\node[data] (codeparrot) {CodeParrot\\(14M C files)};
\node[process, right=of codeparrot] (filter) {Filter to\\self-contained\\(1M files)};
\node[process, right=of filter] (soteria) {Soteria};
\node[data, right=of soteria] (traceconfigs) {Trace\\configurations};
\node[result, right=2.5cm of traceconfigs] (train) {Train\\Qwen3-8B};
\node[result, right=1.5cm of train] (eval) {Benchmark\\(500 tasks)};

\node[data, below=0.4cm of soteria] (conly) {C files only\\(no traces)};
\node[optional, above=0.4cm of traceconfigs] (inform) {Informalize\\(LLM)};

\node[data, above=0.3cm of eval] (svcomp) {SV-COMP 2025};

\draw[arr] (codeparrot) -- (filter);
\draw[arr] (filter) -- (soteria);
\draw[arr] (soteria) -- (traceconfigs);
\draw[arr] (traceconfigs) -- (train);
\draw[arr] (train) -- (eval);
\draw[arr] (svcomp) -- (eval);

\draw[arr] (filter) -- (conly);
\draw[arr] (conly) -| (train);
\draw[optarr] (traceconfigs) -- (inform);
\draw[optarr] (inform) -| node[pos=0.15, above, font=\scriptsize\itshape, text=gray] {optional} (train);

\end{tikzpicture}%
}
\caption{Overview of our data generation and training pipeline. Starting from CodeParrot, we filter to self-contained C files and run Soteria to produce symbolic execution traces. We explore multiple trace configurations (varying format, filtering, and composition; see Appendix~\ref{app:experiment-configs}), with optional informalization. Each configuration is used to train Qwen3-8B and evaluate on our benchmark.}
\label{fig:pipeline}
\end{figure}

\paragraph{Per-Property Patterns.}
The per-property breakdown (Table~\ref{tab:baseline-heat}) shows memory safety violations are hardest to detect (even Claude Opus 4.7 at 86\%), overflow violations show the widest spread (30--100\%), and data race detection varies considerably (37--97\%). Manual review confirms these are genuine reasoning failures, not prompt artifacts (Appendix~\ref{app:failure-example}).

\paragraph{Effect of Program Length.}
Figure~\ref{fig:length-decay-main} shows that holds accuracy remains stable across length bins, while violation detection degrades sharply. For Qwen3-32B, violation accuracy drops from 90.0\% (0--25 LOC) to 32.1\% (101--200 LOC). Even Claude Opus 4.7 dips to 77.7\% on programs of 101--200 lines. Since our benchmark contains only single-file programs under 400 lines, this represents a lower bound on real-world difficulty.

\paragraph{Failure Analysis and Implications.}
An LLM-as-judge analysis of all 1069 false negatives (Appendix~\ref{app:failure-example}) reveals five failure modes: missing the triggering edge case (44\%), identifying the bug but rationalizing it away (27\%), incorrect reasoning (19\%), universal/existential confusion (8\%, concentrated in termination), and superficial analysis (1\%). These diverse patterns suggest no single intervention will address all failure modes.


\section{Data Generation and Training}

Given that training data and objectives appear to matter more than model size for violation detection, we now describe our pipeline for generating training data from symbolic execution artifacts (Figure~\ref{fig:pipeline}).

\subsection{Primary Training Data}

We use Soteria \citep{ayoun2025soteria}, a symbolic execution engine for C and Rust, to generate training data. Unlike concrete execution, which runs a program on specific inputs, symbolic execution reasons over all possible inputs simultaneously, exploring execution paths by treating inputs as symbolic variables and tracking the conditions under which each path is taken. Soteria produces detailed traces that capture memory states, branch conditions, and the reasoning behind property violations, making implicit program semantics explicit. We use only its built-in checks for memory safety, buffer overflows, and undefined behavior, avoiding the need for property synthesis and allowing the pipeline to scale to arbitrary C code without annotation. An example trace is provided in Appendix~\ref{app:soteria-trace}.

As our source corpus, we use the GitHub CodeParrot dataset~\cite{codeparrot}\footnote{CodeParrot is publicly available on Hugging Face. Individual files are subject to their original repository's license. We use this data for research only and do not distribute the model.}, which contains over 14 million C files. We filter to approximately 1 million self-contained C files (no \texttt{\#include} or \texttt{\#import} statements) to keep traces simple. Running Soteria yields approximately 34,000 usable traces. The main bottlenecks are parser limitations (Soteria does not yet support all C variants), files without meaningful control flow, and Soteria's internal path-exploration bounds. Despite this modest volume, our downstream results show meaningful improvement, suggesting room for gains as tooling matures.

This training data is independent of our benchmark: we use arbitrary open-source C code, not SV-COMP programs, allowing us to assess whether improvements transfer to unseen verification tasks. While CodeParrot and SV-COMP both draw from open-source repositories, we did not perform explicit deduplication; even in the unlikely event of overlap, Soteria traces would not reveal ground truth for the benchmark properties (termination, reachability, data races) that Soteria does not check.

\subsection{Training Configurations}

We use Qwen3-8B as our base model for most experiments (with Qwen3-32B as a scale reference), as it is open-weights and tractable for training while still performing reasonably on our baseline evaluation. Our primary training method is continued pretraining (CPT), where traces are concatenated and the model is trained with a standard language modeling objective. We also explore LoRA supervised fine-tuning (SFT) and two-stage CPT+SFT approaches (Appendix~\ref{app:experiment-configs}). To test whether symbolic execution artifacts can improve code understanding, and if so, which aspects matter most, we systematically vary the training data along several dimensions.

\paragraph{Trace Filtering.} Not all Soteria traces are equally informative. Traces fall into three categories: \emph{manifest bugs}, where the symbolic executor proves a violation occurs on all feasible inputs; \emph{latent bugs}, where an error path is found for some inputs but universality is not proven; and \emph{no-violation traces}, where Soteria completes analysis without finding any violation. We systematically vary the training data composition: manifest bugs only (3,208 traces), manifest plus latent bugs (754 manifest + 4,294 latent = 5,048 traces), bug and no-violation traces excluding incomplete analyses (16,234 traces), all traces (34,495 traces), and a balanced 50/50 bug/non-bug split (6,416 traces). We also test trace simplification (stripping to error-relevant lines) and training for multiple epochs.

\paragraph{Trace Format.} Soteria produces interactive HTML reports with collapsible sections, CSS classes for log levels, and HTML-encoded entities (see Appendix~\ref{app:soteria-trace}). We compare training on this HTML output (with non-semantic metadata such as timestamps removed) versus plain-text traces (HTML tags and metadata removed, entities decoded) and informalized traces, where we prompt an LLM to summarize the symbolic execution in natural language (see Appendix~\ref{app:informalized-trace}). We also evaluate the effect of enabling Qwen3's internal chain-of-thought reasoning (``thinking'') at inference time, which appears to benefit strongly from the trace format.

\paragraph{Trace vs.\ Code.} To isolate the contribution of symbolic execution artifacts from the effect of simply training on more C code, we compare training on Soteria traces versus training on the same 3,208 underlying C files alone (without traces), controlling for the number of programs seen.

Training details including hyperparameters and compute resources are provided in Appendix~\ref{app:training-details}.


\section{Results}
\label{sec:results}

\begin{figure}[t]
\centering
\begin{tikzpicture}
\begin{axis}[
    ybar,
    bar width=9pt,
    width=\columnwidth,
    height=3.5cm,
    ylabel={Accuracy (\%)},
    ymin=0, ymax=115,
    xtick={1,2,3,4,5,6,7,8,9},
    xticklabels={
        {8B base\\(no think)},
        {+ C files\\(no think)},
        {+ all HTML\\(no think)},
        {+ bug HTML\\(no think)},
        {+ bug plain\\(no think)},
        {8B base\\(think)},
        {+ bug HTML\\(think)},
        {32B base\\(no think)},
        {32B base\\(think)}
    },
    xticklabel style={font=\tiny, align=center, rotate=45, anchor=east},
    legend style={at={(0.5,1.05)}, anchor=south, font=\small, legend columns=3},
    enlarge x limits=0.08,
    grid=major,
    grid style={dashed, gray!30},
    xmin=0.5, xmax=9.5,
]
\addplot[bar width=12pt, bar shift=0, fill=blue!50, draw=blue!70,
    nodes near coords,
    nodes near coords style={font=\fontsize{5}{6}\selectfont, anchor=south},
    point meta=explicit symbolic,
    error bars/.cd, y dir=both, y explicit, error bar style={line width=0.4pt, color=blue!70}, error mark options={rotate=90, mark size=1.5pt, color=blue!70}] coordinates {(1,90.7) +- (0,0.6) [70.0] (2,87.4) +- (0,0.4) [63.8] (3,83.8) +- (0,1.8) [61.8] (4,80.0) +- (0,0.7) [68.3] (5,71.6) +- (0,2.9) [62.6] (6,85.4) +- (0,1.7) [66.7] (7,78.0) +- (0,2.1) [72.7] (8,95.4) +- (0,0.6) [76.2] (9,95.2) +- (0,0.3) [86.5]
};
\addplot[bar width=12pt, bar shift=0, fill=green!50, draw=green!70] coordinates {(1,70.0) (2,63.8) (3,61.8) (4,68.3) (5,62.6) (6,66.7) (7,72.7) (8,76.2) (9,86.5)
};
\addplot[bar width=12pt, bar shift=0, fill=orange!60, draw=orange!80,
    error bars/.cd, y dir=both, y explicit, error bar style={line width=0.4pt, color=orange!80}, error mark options={rotate=90, mark size=1.5pt, color=orange!80}] coordinates {(1,49.4) +- (0,2.1) (2,40.2) +- (0,2.7) (3,39.9) +- (0,2.3) (4,56.7) +- (0,1.0) (5,53.7) +- (0,0.3) (6,48.0) +- (0,2.0) (7,67.3) +- (0,1.6) (8,56.9) +- (0,1.5) (9,77.8) +- (0,0.7)
};
\legend{Property Holds, Overall, Property Violated}
\end{axis}
\end{tikzpicture}
\caption{Accuracy on our benchmark by training configuration, split by property holds and violated. Numbers above each bar show overall accuracy (\%). All 8B models use Qwen3-8B as the base model; ``think'' denotes inference with chain-of-thought reasoning enabled. Bug traces with thinking achieve the best 8B violation detection (67.3\%), outperforming the 32B base model on violations (56.9\%).}
\label{fig:ablation-bar}
\end{figure}
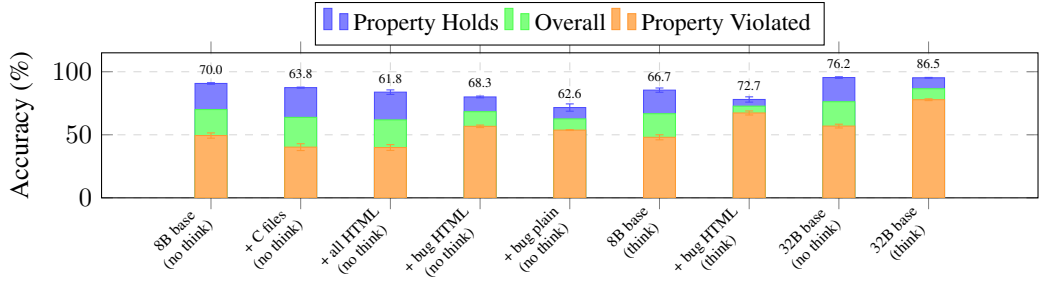

The results of training Qwen3-8B in the configurations described above are shown in Figure~\ref{fig:ablation-bar} (key patterns) and Table~\ref{tab:heat} (per-property breakdowns). The clearest finding is that continued pretraining on Soteria's manifest bug traces in HTML format, combined with thinking at inference time, yields the strongest improvement in violation detection (+17.9 pp vs.\ the no-thinking baseline, $p{<}0.01$ paired $t$-test across 4 runs; per-task McNemar's test confirms significance across the task distribution: $p{<}10^{-7}$) and a net gain in overall accuracy (+2.7 pp, $p{<}0.05$), while other configurations either underperform or introduce trade-offs.

\paragraph{Bug Traces and Trace Curation.}
CPT on 3,208 HTML-formatted manifest bug traces increases violation detection from 49.4\% to 56.7\% (+7.3 pp) without thinking. With thinking, CPT improves violation detection from 48.0\% to 67.3\% (+19.3 pp). The synergy is superadditive: thinking alone provides no benefit ($-$1.4 pp), but combined with CPT adds +10.6 pp beyond CPT alone.

Training on all 34,495 HTML traces (including latent bugs and no-violation traces) hurts violation detection (39.9\%), as does training on C source files alone (40.2\%), both \emph{below} the 49.4\% baseline. This confirms that semantic signal in Soteria's bug traces, not code volume, drives the improvement.

\paragraph{Dataset Size.}
The best CPT dataset consists of just 3,208 traces (${\sim}$4.6M tokens), trained for a single epoch (${\sim}$10 minutes on 8 GPUs). Larger datasets consistently performed worse: 5,048 traces yielded 50.2\% violated, 16,234 traces yielded 47.5\%, and 34,495 traces yielded 39.9\%, all below the 56.7\% achieved by 3,208 traces. The expanded and filtered datasets also differ in bug composition (the 5,048-trace set subsamples manifest bugs to 754), so the size effect is partially confounded with curation; the consistent pattern nonetheless suggests signal concentration matters more than quantity.

\paragraph{Trace Format.}
Preserving Soteria's HTML markup yields better results than plain text. On 3,208 bug traces without thinking, HTML achieves 56.7\% violated versus 53.7\% for plain text ($-$3.0 pp). However, per-property differences are large: plain text outperforms HTML on overflow (68.0\% vs.\ 51.0\%) and data race (73.0\% vs.\ 65.5\%), while being far worse on memory safety (22.0\% vs.\ 39.0\%), suggesting the relationship between format and performance may be property-dependent. With thinking, the gap widens dramatically: HTML reaches 67.3\% while plain text collapses to 33.2\% with a 45\% unparsed rate. Informalized traces (LLM-generated natural-language summaries) collapsed to 25.2\% overall with 62\% unparsed (Appendix~\ref{app:experiment-configs}), confirming that structured tool output matters, not just semantic content.

\paragraph{Per-Property Patterns.}
Table~\ref{tab:heat} shows that the best configuration (HTML bug traces + thinking) improves violation detection across all five properties compared to the base model without thinking. Overflow sees the largest gain (+28.5 pp to 66.0\%, $p{<}0.001$), followed by memory safety (+28.0 pp to 47.0\%, $p{<}0.01$), termination (+16.0 pp to 94.5\%, $p{<}0.01$), data races (+11.5 pp to 69.0\%), and reachability (+5.5 pp to 60.0\%). The data race and reachability improvements do not reach significance at $p{<}0.05$ with 4 runs and 50 tasks per cell, though both show consistent positive trends. Memory safety remains the hardest property (47.0\%).

\begin{table}[t]
\centering
\scriptsize
\renewcommand{\arraystretch}{0.65}
\caption{Per-property accuracy (\%) for selected configurations. Full results in Appendix~\ref{app:training-results-detailed}.}
\label{tab:heat}
\resizebox{\columnwidth}{!}{%
\begin{tabular}{p{3.8cm}|c|c|c|c|c|c}
\multicolumn{7}{c}{\textit{Property Holds}} \\
\textbf{Configuration} & \textbf{Total} & \textbf{Mem. Safety} & \textbf{Overflow} & \textbf{Termin.} & \textbf{Reach.} & \textbf{Data Race} \\
\hline
8B base (no thinking) & \heatcellpm{90.7}{0.6} & \heatcellpm{96.0}{1.4} & \heatcellpm{89.0}{2.2} & \heatcellpm{96.0}{1.4} & \heatcellpm{89.5}{4.6} & \heatcellpm{83.0}{5.0} \\
\quad + C files only & \heatcellpm{87.4}{0.4} & \heatcellpm{94.0}{2.4} & \heatcellpm{91.0}{3.3} & \heatcellpm{96.5}{0.9} & \heatcellpm{83.5}{3.3} & \heatcellpm{72.0}{2.4} \\
\quad + All traces (HTML) & \heatcellpm{83.8}{1.8} & \heatcellpm{83.0}{4.1} & \heatcellpm{88.5}{1.7} & \heatcellpm{94.5}{2.6} & \heatcellpm{83.0}{1.0} & \heatcellpm{70.0}{7.5} \\
\quad + Bug traces (HTML) & \heatcellpm{80.0}{0.7} & \heatcellpm{93.0}{1.7} & \heatcellpm{79.0}{1.0} & \heatcellpm{93.5}{1.7} & \heatcellpm{77.0}{2.2} & \heatcellpm{57.5}{4.3} \\
\quad + Bug traces (plain text) & \heatcellpm{71.6}{2.9} & \heatcellpm{81.5}{1.7} & \heatcellpm{60.0}{5.1} & \heatcellpm{88.0}{3.5} & \heatcellpm{76.0}{6.8} & \heatcellpm{52.5}{5.9} \\
\hline
8B base (thinking) & \heatcellpm{85.4}{1.7} & \heatcellpm{95.5}{1.7} & \heatcellpm{79.0}{3.3} & \heatcellpm{94.5}{1.7} & \heatcellpm{81.5}{1.7} & \heatcellpm{76.5}{3.8} \\
\quad + Bug traces (HTML) & \heatcellpm{78.0}{2.1} & \heatcellpm{79.5}{3.0} & \heatcellpm{71.0}{4.1} & \heatcellpm{83.0}{2.2} & \heatcellpm{80.5}{3.6} & \heatcellpm{76.0}{5.1} \\
\hline
32B base (no thinking) & \heatcellpm{95.4}{0.6} & \heatcellpm{97.5}{0.9} & \heatcellpm{94.0}{3.7} & \heatcellpm{94.5}{0.9} & \heatcellpm{94.0}{2.4} & \heatcellpm{97.0}{1.0} \\
32B base (thinking) & \heatcellpm{95.2}{0.3} & \heatcell{98.0} & \heatcellpm{93.0}{1.7} & \heatcellpm{92.5}{2.2} & \heatcellpm{93.5}{0.9} & \heatcellpm{99.0}{1.0} \\
\hline
\multicolumn{7}{c}{\textit{Property Violated}} \\
\textbf{Configuration} & \textbf{Total} & \textbf{Mem. Safety} & \textbf{Overflow} & \textbf{Termin.} & \textbf{Reach.} & \textbf{Data Race} \\
\hline
8B base (no thinking) & \heatcellpm{49.4}{2.1} & \heatcellpm{19.0}{3.6} & \heatcellpm{37.5}{1.7} & \heatcellpm{78.5}{1.7} & \heatcellpm{54.5}{5.2} & \heatcellpm{57.5}{1.7} \\
\quad + C files only & \heatcellpm{40.2}{2.7} & \heatcellpm{22.5}{3.8} & \heatcellpm{33.0}{3.3} & \heatcellpm{59.0}{7.3} & \heatcellpm{43.0}{2.2} & \heatcellpm{43.5}{1.7} \\
\quad + All traces (HTML) & \heatcellpm{39.9}{2.3} & \heatcellpm{20.5}{4.8} & \heatcellpm{25.5}{4.1} & \heatcellpm{65.5}{3.0} & \heatcellpm{32.0}{3.7} & \heatcellpm{56.0}{4.9} \\
\quad + Bug traces (HTML) & \heatcellpm{56.7}{1.0} & \heatcellpm{39.0}{4.1} & \heatcellpm{51.0}{3.0} & \heatcellpm{79.5}{1.7} & \heatcellpm{48.5}{3.0} & \heatcellpm{65.5}{3.0} \\
\quad + Bug traces (plain text) & \heatcellpm{53.7}{0.3} & \heatcellpm{22.0}{1.4} & \heatcellpm{68.0}{3.7} & \heatcellpm{66.0}{4.0} & \heatcellpm{39.5}{4.6} & \heatcellpm{73.0}{3.0} \\
\hline
8B base (thinking) & \heatcellpm{48.0}{2.0} & \heatcellpm{24.5}{2.2} & \heatcellpm{44.5}{4.6} & \heatcellpm{82.5}{3.3} & \heatcellpm{43.5}{3.0} & \heatcellpm{45.0}{4.1} \\
\quad + Bug traces (HTML) & \heatcellpm{67.3}{1.6} & \heatcellpm{47.0}{1.7} & \heatcellpm{66.0}{2.4} & \heatcellpm{94.5}{1.7} & \heatcellpm{60.0}{1.4} & \heatcellpm{69.0}{8.5} \\
\hline
32B base (no thinking) & \heatcellpm{56.9}{1.5} & \heatcellpm{39.0}{4.4} & \heatcellpm{52.5}{3.6} & \heatcellpm{88.5}{1.7} & \heatcellpm{55.0}{4.1} & \heatcellpm{49.5}{2.2} \\
32B base (thinking) & \heatcellpm{77.8}{0.7} & \heatcellpm{69.0}{3.0} & \heatcellpm{72.5}{1.7} & \heatcellpm{95.5}{1.7} & \heatcellpm{77.5}{0.9} & \heatcellpm{74.5}{1.7} \\
\end{tabular}%
}
\end{table}

\paragraph{Scale Comparison.}
The best 8B configuration (72.7\% overall) approaches the Qwen3-32B base model without thinking (76.2\%) while substantially outperforming it on violation detection (67.3\% vs.\ 56.9\%). The 32B model with thinking reaches 86.5\% overall, indicating significant headroom from scaling that CPT at 8B does not close. CPT on the 32B model is a promising direction given the strong CPT--thinking synergy at 8B scale. Figure~\ref{fig:ablation-length} in the appendix shows that the CPT improvement over the base model holds across all program length bins, though the length-decay pattern persists.

\paragraph{Thinking and Output Budget.}
Thinking mode increases unevaluated response rates (9--45\%): the model's chain-of-thought reasoning frequently exhausts the output token budget before producing a final answer, yielding responses that cannot be parsed into a verdict. The best configuration leaves 13.4\% of tasks unevaluated (vs.\ 1.6\% for the base model), so the fixed-denominator numbers (counting unevaluated responses as incorrect) throughout this paper are conservative. Crucially, the unevaluated rate is nearly symmetric across ground truth classes (14.1\% on holds vs.\ 12.7\% on violated), ruling out the hypothesis that the model selectively times out on safe programs while searching for nonexistent bugs. On parseable responses only, the best configuration achieves 90.8\% holds and 77.1\% violated (83.9\% overall), compared to 91.2\% holds and 50.8\% violated (71.2\% overall) for the base model --- a +26 pp violated improvement with no degradation on holds. Re-evaluating with 3$\times$ the output tokens (24,576 vs.\ 8,192) improved overall accuracy by +1.8 pp, but 10--14\% of tasks still time out, suggesting the bottleneck shifts to compute time.

\paragraph{Additional Findings.}
We explored 28 configurations in total (Appendix~\ref{app:training-results-detailed}). LoRA SFT exhibited a sharp capability--stability tradeoff, stripping traces to error-relevant lines degraded performance, and adding an SFT stage after CPT hurt violation detection. As a prompting baseline, we evaluated the base model with two Soteria trace examples (one showing a bug, one showing safe execution) included in the prompt. This few-shot approach hurt violation detection ($-$8.3 pp to 41.1\%) compared to the zero-shot base model, degrading all five properties. The model cannot acquire verification reasoning from in-context examples; CPT is necessary to internalize these patterns.


\section{Discussion}
\label{sec:discussion}

\subsection{Interpreting the Results}

The superadditive synergy between CPT and thinking (Section~\ref{sec:results}) suggests that CPT instills latent verification knowledge the model cannot express without extended reasoning. CPT operates through two channels: it improves reasoning accuracy on the responses the model produces, and it makes chain-of-thought more structured, reducing the fraction of tasks where the model exhausts its output budget. CPT teaches \emph{what} to reason about, while thinking provides the budget to \emph{apply} that reasoning.

The advantage of HTML traces over plain text likely originates from the explicit hierarchical information (nested tags, log-level annotations) providing anchoring points for multi-step reasoning. The holds--violated gap narrows after training. Memory safety remains the hardest property; Appendix~\ref{app:cpt-qualitative} shows a qualitative example illustrating the pointer reasoning that training enables.

\subsection{Generality and Scalability}

Our approach requires no labeled data, no task-specific prompts during training, and no manual modification to the tool's output format, so it can be applied to any formal analysis tool producing structured textual output (bounded model checkers, abstract interpreters, fuzz testing harnesses). Training cost is minimal (${\sim}$10 minutes on 8 GPUs for 3,208 traces).

Notably, Soteria-only training improved violation detection across all five benchmark properties, including data races and reachability, which Soteria does not specifically analyze. Soteria's built-in checks cover memory safety, buffer overflows, and undefined behavior, so the training traces overlap with the benchmark's memory safety and overflow categories. However, comparing CPT against the base model both with thinking enabled, gains extend to termination (+12.0 pp), data races (+24.0 pp), and reachability (+16.5 pp), suggesting the model learns general verification reasoning patterns.

\subsection{Limitations}
\label{sec:limitations}

Our experiments use a single base model (Qwen3-8B, with Qwen3-32B as a scale reference), a single symbolic execution engine (Soteria), and a single language (C). We evaluate only binary verdicts on single-file programs; evaluating machine-checkable witnesses or proofs would require substantial additional work. We have not assessed transfer to downstream tasks. Per-property comparisons have limited statistical power with 50 tasks per cell and 4 runs. We have not tested for benchmark contamination, though the ground truth verdicts (whether each property holds or is violated) are established by competition verification tools and are unlikely to appear in pretraining corpora alongside the programs. Each limitation represents a concrete avenue for future work.

\subsection{Broader Impact}
\label{sec:broader-impact}

Improving LLMs' ability to detect property violations could benefit software reliability by catching bugs earlier, but could also lower the barrier to automated vulnerability discovery if scaled to real-world codebases. We do not release trained model weights.


\section{Conclusion}

We presented an approach for improving LLM understanding of low-level code by training on symbolic execution artifacts, alongside a benchmark of 500 C verification tasks from SV-COMP 2025. Our baseline evaluation of 14 models revealed that violation detection is consistently harder than confirming properties hold, independent of model size.

By training Qwen3-8B on just 3,208 Soteria bug traces with chain-of-thought reasoning at inference time, we achieved a 17.9 pp improvement on violation detection. On parseable responses, accuracy on the property holds case is preserved (90.8\%) and violation detection improves by +26 pp, confirming genuine reasoning improvement. Ablations confirm that trace semantics, not code volume, drive the improvement. The pipeline requires no annotation and scales with open-source code availability.

Future work includes extending to multi-file programs, incorporating additional formal tools and languages, and using symbolic execution in a reinforcement learning loop.


\begin{ack}
This work was funded by Amazon Web Services. We thank Sacha-Élie Ayoun, Azalea Raad, and Opale Sjöstedt for their help with Soteria.
\end{ack}



\appendix

\section{CBMC Regression Tests}
\label{app:cbmc}

We initially evaluated models on the CBMC regression tests \citep{cbmc-regression}, a collection of approximately 1,000 self-contained C files where ground truth is established by the CBMC verifier. We constructed prompts to have LLMs act as static analyzers, producing binary judgments (property holds or violated). All models were evaluated via a commercial inference API.

Manual inspection of model responses revealed several systematic issues with the evaluation. Some test cases involved uninitialized variables, strict aliasing violations, or behavior dependent on system specifics, compiler version, or C standard interpretation, making ground truth ambiguous. In other cases, CBMC command-line options contained information crucial to the verdict that was not provided to the model. We filtered out these problematic cases, reducing the evaluation set to 179 tasks with unambiguous ground truth. Among the remaining model errors, we observed failures in basic arithmetic, missing knowledge about C standard library semantics or memory layout, off-by-one errors in array index reasoning, and misinterpretation of compiler warnings as runtime errors.

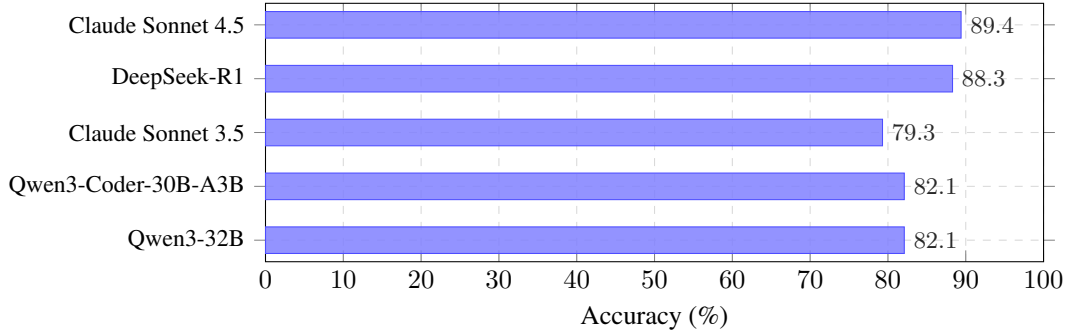
\begin{figure}[H]
\centering
\begin{tikzpicture}
\begin{axis}[
    xbar,
    bar width=10pt,
    width=0.85\columnwidth,
    height=5cm,
    xlabel={Accuracy (\%)},
    xmin=0, xmax=100,
    symbolic y coords={Qwen3-32B,Qwen3-Coder-30B-A3B,Claude Sonnet 3.5,DeepSeek-R1,Claude Sonnet 4.5},
    ytick=data,
    yticklabel style={font=\small},
    nodes near coords,
    nodes near coords style={font=\small, anchor=west},
    grid=major,
    grid style={dashed, gray!30},
    every axis plot/.append style={fill opacity=0.85},
]
\addplot[fill=blue!50, draw=blue!70] coordinates {
(82.1,Qwen3-32B)
(82.1,Qwen3-Coder-30B-A3B)
(79.3,Claude Sonnet 3.5)
(88.3,DeepSeek-R1)
(89.4,Claude Sonnet 4.5)
};
\end{axis}
\end{tikzpicture}
\caption{Model performance on CBMC regression tests (\% correct out of 179 tasks). High accuracy across models indicates the benchmark is too easy to differentiate improvements.}
\label{tab:cbmc-results}
\end{figure}

As shown in Figure~\ref{tab:cbmc-results}, all models achieve 79--89\% accuracy on the filtered set, with accuracy too clustered to meaningfully differentiate improvements. This motivated the construction of our benchmark.

\subsection{Prompt Template}
\label{app:cbmc-prompt}

The following prompt template was used for evaluation on CBMC regression tests:

\begin{quote}
\itshape
You are an expert in programming with C and are asked to analyse the code below. Your task is to verify that this code would run correctly without producing the following types of runtime errors:
\begin{itemize}
\item Buffer overflows
\item Array bounds violations
\item Pointer safety
\item Memory leaks
\item Division by zero
\item Assertion violations
\end{itemize}

Please go through the code carefully step-by-step and consider which conditions hold throughout its execution. You should keep track of variables and pointers and at every step of the execution think about whether a runtime error will occur.

\{c\_code\}

Please end your answer with the precise string ``Final answer: ERROR'' if you identified runtime errors or ``Final answer: OKAY'' if there are no errors.
\end{quote}

\subsection{Results}

\label{app:cbmc-results}

\section{Benchmark and Baseline Evaluation}
\label{app:svcomp}

\subsection{Statistics}
\label{app:svcomp-stats}

\begin{figure}[H]
\centering
\begin{tikzpicture}
\begin{axis}[
    ybar,
    bar width=20pt,
    width=0.85\columnwidth,
    height=5cm,
    ylabel={Number of programs},
    xlabel={Lines of code},
    ymin=0,
    xtick={1,2,3,4,5},
    xticklabels={0--25,26--50,51--100,101--200,201--400},
    xticklabel style={font=\small},
    nodes near coords,
    nodes near coords style={font=\small, anchor=south},
    grid=major,
    grid style={dashed, gray!30},
    every axis plot/.append style={fill opacity=0.85},
    xmin=0.5, xmax=5.5,
]
\addplot[fill=blue!50, draw=blue!70] coordinates {
(1,49) (2,157) (3,186) (4,50) (5,46)
};
\end{axis}
\end{tikzpicture}
\caption{Distribution of program lengths in our benchmark (488 unique programs across 500 tasks, median 56 LOC, range 8--398).}
\label{fig:svcomp-length-dist}
\end{figure}
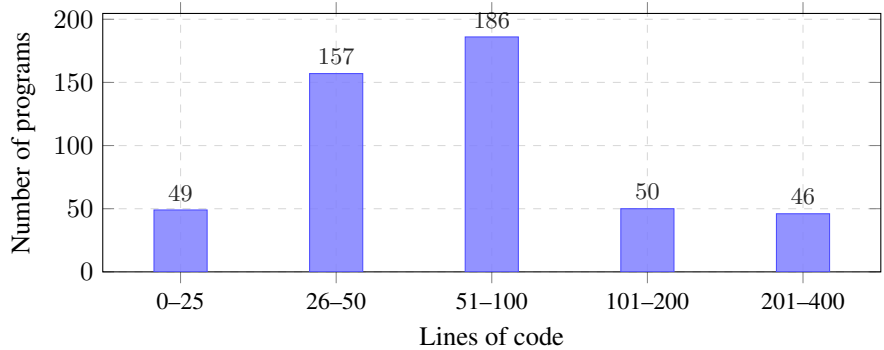

Figure~\ref{fig:svcomp-length-dist} shows the distribution of program lengths in our benchmark.

\subsection{Preprocessing}
\label{app:svcomp-preprocessing}

The original SV-COMP 2025 benchmark was not designed for LLM evaluation. We apply preprocessing steps including the following (see Section~\ref{sec:benchmark} for additional curation such as balancing the dataset):
\begin{itemize}
\item Exclude files with incorrect specifications (from InvalidTasks-Def file)
\item Exclude additional invalid files identified by SV-COMP validation scripts
\item Exclude tasks referencing multiple C files
\item Exclude tasks with malformed YAML metadata
\item Exclude files with \texttt{\#include} of custom headers
\item Remove comments from C code
\item Adjust prompts for correct treatment of uninitialized memory
\item Adjust prompts to include SV-COMP specification details (\texttt{\_\_VERIFIER\_atomic\_begin/end})
\item Ensure model returns "ERROR" only for violations of the specified property
\end{itemize}

\subsection{Evaluation Prompt}
\label{app:svcomp-prompt}

For evaluation on our benchmark, we use a prompt with property-specific descriptions and instructions.

\subsubsection{Prompt Template}

\begin{quote}
\itshape
You are an expert in programming with C and are asked to analyse the code below. Your task is to verify that this code would run correctly \{property\_description\}

Please go through the code carefully step-by-step and consider which conditions hold throughout its execution. You should keep track of variables and pointers and at every step of the execution think about whether an error will occur. Note that this program uses the \{data\_model\} data model.

\textbf{Note:}
\begin{itemize}
\item The code may contain functions like \texttt{\_\_VERIFIER\_nondet\_X()}. You should assume that this returns an arbitrary value of the indicated type (no side effects). You can assume that these functions are implemented according to the following template: \texttt{X \_\_VERIFIER\_nondet\_X() \{\{ X val; return val; \}\}}
\item \texttt{\_\_VERIFIER\_atomic\_begin()} and \texttt{\_\_VERIFIER\_atomic\_end()}: In a multi-threaded run-time environment, if a list of statements is placed between two function calls you can assume that they are executed atomically. That is, you can assume that the execution between those calls is not interrupted.
\item \texttt{malloc()}, \texttt{free()}: Assume that the functions malloc and alloca always return a valid pointer, i.e., the memory allocation never fails, and function free always deallocates the memory and makes the pointer invalid for further dereferences.
\end{itemize}

\{c\_code\}

\{property\_instruction\}
\end{quote}

\noindent The \texttt{\{data\_model\}} placeholder is populated with the task's data model from SV-COMP metadata (ILP32 for 397 tasks, LP64 for 91 tasks), which determines integer and pointer sizes.

\subsubsection{Property Descriptions}

\textbf{No Overflow:}
\begin{quote}
\itshape
without producing an overflow error. For the context of this task, \textbf{overflow error} means the following: It can never happen that the resulting type of an operation is a signed-integer type but the resulting value is not in the range of values that are representable by that type. A violation of this property matches what C11 defines as undefined behavior. (Hence, conversions to signed-integer types do not violate this property.)
\end{quote}

\textbf{Memory Safety:}
\begin{quote}
\itshape
without memory safety violations. This means that the following three conditions have to hold: 
1. Valid Deallocating: All memory deallocations are valid (counterexample: invalid free). More precisely: There exists no execution of the program on which an invalid memory deallocation occurs. 
2. Valid Dereferencing: All pointer dereferences are valid (counterexample: invalid dereference). More precisely: There exists no execution of the program on which an invalid pointer dereference occurs.
3. Valid Memory Tracking: All allocated memory blocks are tracked. The set of tracked blocks is defined as the smallest set of blocks satisfying the following two rules:
A block is tracked whenever there is a pointer to this block (not necessarily pointing to the beginning of the block) or to the first address after this block (see 6.5.6 of C11 standard) stored in a program variable. The variable can be of a pointer type or of a compound type containing a pointer. The variable does not have to be in the current scope, it can be global or on the call stack.
If some pointer in a tracked block points to another block (again, not necessarily to the beginning of the block) or to the first address after this block, this pointed block is also tracked. In particular, a leaked memory block is not tracked. Hence, a program with a memory leak does not satisfy this property.
Note that reading uninitialised memory is considered okay in the context of this task. You are meant to check issues with (de)allocation, but not initialisation.
\end{quote}

\textbf{Termination:}
\begin{quote}
\itshape
and will always terminate. In the context of this task, \textbf{termination} means that every execution path finally reaches the end of the program.
\end{quote}

\textbf{Unreachable Call:}
\begin{quote}
\itshape
without calling the error function. This means that the function \texttt{reach\_error} is not called in any finite execution of the program.
\end{quote}

\textbf{No Data Race:}
\begin{quote}
\itshape
without any data races occurring. For the context of this task, \textbf{data race} means the following: If there exist two or more concurrent accesses to the same memory location and at least one is a write access, then all accesses must be atomic.
\end{quote}

\subsubsection{Property Instructions}

\textbf{No Overflow:}
\begin{quote}
\itshape
Please end your answer with the precise string ``Final answer: OKAY'' if the code does not run into overflow errors. Otherwise, you should end your answer with ``Final answer: ERROR''. Even if there are problems with the code, as long as no overflow error occurs you should respond with ``Final Answer: OKAY''.
\end{quote}

\textbf{Memory Safety:}
\begin{quote}
\itshape
Please end your answer with the precise string ``Final answer: OKAY'' if the code is memory safe. Otherwise, you should end your answer with ``Final answer: ERROR''. Even if there are problems with the code, as long as none of the mentioned memory safety issues occurs you should respond with ``Final Answer: OKAY''. Remember that the focus is on (de)allocation and that accessing uninitialised memory is not considered a problem in the context of this task.
\end{quote}

\textbf{Termination:}
\begin{quote}
\itshape
Please end your answer with the precise string ``Final answer: OKAY'' if the program terminates. Otherwise, you should end your answer with ``Final answer: ERROR''. Even if there are problems with the code, as long as it terminates you should respond with ``Final Answer: OKAY''.
\end{quote}

\textbf{Unreachable Call:}
\begin{quote}
\itshape
Please end your answer with the precise string ``Final answer: OKAY'' if the error function \texttt{reach\_error} is not called. Otherwise, you should end your answer with ``Final answer: ERROR''. Even if there are problems with the code, as long as the error function is not called you should respond with ``Final Answer: OKAY''.
\end{quote}

\textbf{No Data Race:}
\begin{quote}
\itshape
Please end your answer with the precise string ``Final answer: OKAY'' if there are no data races. Otherwise, you should end your answer with ``Final answer: ERROR''. Even if there are problems with the code, as long as there is no data race you should respond with ``Final Answer: OKAY''.
\end{quote}

\subsection{Detailed Baseline Results}
\label{app:svllm-baseline-detailed}

Aggregate accuracy is shown in Figure~\ref{fig:baseline-bar} in the main text.

\begin{table}[H]
\centering
\scriptsize
\vspace{-1em}
\renewcommand{\arraystretch}{0.60}
\caption{Baseline per-property accuracy (\%) on our benchmark across all program lengths. Top: property holds; bottom: property violated.}
\label{tab:baseline-heat}
\resizebox{\columnwidth}{!}{%
\begin{tabular}{l|c|c|c|c|c|c}
\multicolumn{7}{c}{\textit{Property Holds (1--400 LOC)}} \\
\textbf{Model} & \textbf{Total} & \textbf{Mem. Safety} & \textbf{Overflow} & \textbf{Termin.} & \textbf{Reach.} & \textbf{Data Race} \\
\hline
Claude Opus 4.7 & \heatcellpm{98.3}{0.2} & \heatcell{100.0} & \heatcellpm{93.5}{0.9} & \heatcellpm{98.5}{0.9} & \heatcell{100.0} & \heatcellpm{99.5}{0.9} \\
Mistral Large 3 675B & \heatcellpm{97.8}{0.4} & \heatcellpm{98.0}{1.4} & \heatcell{96.0} & \heatcell{100.0} & \heatcellpm{98.5}{0.9} & \heatcellpm{96.5}{2.6} \\
Kimi K2.5 & \heatcellpm{97.7}{0.8} & \heatcellpm{98.5}{1.7} & \heatcellpm{93.0}{1.0} & \heatcellpm{99.0}{1.0} & \heatcell{100.0} & \heatcellpm{98.0}{1.4} \\
Qwen3-Coder-30B & \heatcellpm{97.5}{1.1} & \heatcellpm{97.0}{1.7} & \heatcellpm{96.0}{1.4} & \heatcellpm{99.0}{1.0} & \heatcellpm{98.5}{0.9} & \heatcellpm{97.0}{2.2} \\
Devstral 2 123B & \heatcellpm{97.2}{0.3} & \heatcellpm{98.5}{0.9} & \heatcellpm{94.0}{1.4} & \heatcellpm{99.0}{1.0} & \heatcellpm{99.5}{0.9} & \heatcellpm{95.0}{1.0} \\
DeepSeek R1 & \heatcellpm{93.7}{0.7} & \heatcellpm{93.5}{1.7} & \heatcellpm{88.0}{2.4} & \heatcellpm{97.5}{1.7} & \heatcellpm{94.0}{2.4} & \heatcellpm{95.5}{3.0} \\
DeepSeek V3.2 & \heatcellpm{95.8}{0.4} & \heatcellpm{96.5}{0.9} & \heatcellpm{91.5}{1.7} & \heatcellpm{97.5}{1.7} & \heatcellpm{96.5}{0.9} & \heatcellpm{97.0}{1.0} \\
Qwen3-32B & \heatcellpm{95.4}{0.6} & \heatcellpm{97.5}{0.9} & \heatcellpm{94.0}{3.7} & \heatcellpm{94.5}{0.9} & \heatcellpm{94.0}{2.4} & \heatcellpm{97.0}{1.0} \\
Claude Sonnet 4.5 & \heatcellpm{94.9}{1.0} & \heatcellpm{94.5}{1.7} & \heatcellpm{90.5}{3.0} & \heatcellpm{98.0}{1.4} & \heatcellpm{99.5}{0.9} & \heatcellpm{92.0}{1.4} \\
Ministral 3 8B & \heatcellpm{93.8}{0.8} & \heatcellpm{96.5}{1.7} & \heatcellpm{91.0}{3.0} & \heatcellpm{99.0}{1.0} & \heatcellpm{91.0}{2.2} & \heatcellpm{91.5}{2.2} \\
Qwen3-Coder-480B & \heatcellpm{93.7}{1.1} & \heatcellpm{92.0}{1.4} & \heatcellpm{88.0}{4.5} & \heatcellpm{97.5}{0.9} & \heatcellpm{97.0}{1.0} & \heatcellpm{94.0}{2.0} \\
GPT-OSS 120B & \heatcellpm{91.9}{0.9} & \heatcellpm{99.0}{1.7} & \heatcellpm{92.0}{2.4} & \heatcellpm{76.5}{1.7} & \heatcellpm{93.0}{1.0} & \heatcellpm{99.0}{1.0} \\
Qwen3-8B & \heatcellpm{90.7}{0.6} & \heatcellpm{96.0}{1.4} & \heatcellpm{89.0}{2.2} & \heatcellpm{96.0}{1.4} & \heatcellpm{89.5}{4.6} & \heatcellpm{83.0}{5.0} \\
GPT-OSS 20B & \heatcellpm{84.8}{1.0} & \heatcellpm{93.0}{3.0} & \heatcellpm{73.5}{2.6} & \heatcellpm{81.0}{5.4} & \heatcellpm{82.5}{4.6} & \heatcellpm{94.0}{1.4} \\
\end{tabular}%
}
\end{table}

\begin{table}[H]
\centering
\scriptsize
\renewcommand{\arraystretch}{0.60}
\resizebox{\columnwidth}{!}{%
\begin{tabular}{l|c|c|c|c|c|c}
\multicolumn{7}{c}{\textit{Property Violated (1--400 LOC)}} \\
\textbf{Model} & \textbf{Total} & \textbf{Mem. Safety} & \textbf{Overflow} & \textbf{Termin.} & \textbf{Reach.} & \textbf{Data Race} \\
\hline
Claude Opus 4.7 & \heatcellpm{92.0}{0.6} & \heatcellpm{85.5}{1.7} & \heatcell{100.0} & \heatcellpm{89.0}{3.6} & \heatcellpm{89.0}{1.0} & \heatcellpm{96.5}{1.7} \\
GPT-OSS 120B & \heatcellpm{90.8}{0.6} & \heatcellpm{88.5}{1.7} & \heatcellpm{92.0}{1.4} & \heatcell{100.0} & \heatcellpm{84.0}{1.4} & \heatcellpm{89.5}{3.3} \\
DeepSeek V3.2 & \heatcellpm{87.8}{1.9} & \heatcellpm{83.5}{5.7} & \heatcellpm{87.5}{0.9} & \heatcellpm{91.5}{3.6} & \heatcellpm{84.5}{3.0} & \heatcellpm{92.0}{2.4} \\
DeepSeek R1 & \heatcellpm{85.2}{1.3} & \heatcellpm{81.5}{5.9} & \heatcellpm{78.0}{3.7} & \heatcellpm{94.0}{1.4} & \heatcellpm{82.0}{2.4} & \heatcellpm{90.5}{2.2} \\
Kimi K2.5 & \heatcellpm{86.1}{1.1} & \heatcellpm{85.0}{3.6} & \heatcellpm{84.5}{1.7} & \heatcellpm{94.0}{1.4} & \heatcellpm{73.5}{1.7} & \heatcellpm{93.5}{3.0} \\
Claude Sonnet 4.5 & \heatcellpm{84.2}{0.7} & \heatcellpm{86.5}{0.9} & \heatcellpm{85.5}{2.2} & \heatcellpm{85.0}{2.2} & \heatcellpm{67.5}{2.6} & \heatcellpm{96.5}{1.7} \\
GPT-OSS 20B & \heatcellpm{84.1}{1.3} & \heatcellpm{81.5}{3.0} & \heatcellpm{81.5}{1.7} & \heatcellpm{97.5}{2.2} & \heatcellpm{79.0}{1.0} & \heatcellpm{81.0}{4.1} \\
Qwen3-Coder-480B & \heatcellpm{74.8}{1.5} & \heatcellpm{73.0}{2.2} & \heatcellpm{81.5}{2.2} & \heatcellpm{81.5}{3.0} & \heatcellpm{64.0}{2.4} & \heatcellpm{74.0}{3.5} \\
Qwen3-32B & \heatcellpm{56.9}{1.5} & \heatcellpm{39.0}{4.4} & \heatcellpm{52.5}{3.6} & \heatcellpm{88.5}{1.7} & \heatcellpm{55.0}{4.1} & \heatcellpm{49.5}{2.2} \\
Mistral Large 3 675B & \heatcellpm{50.2}{1.3} & \heatcellpm{47.0}{4.6} & \heatcellpm{50.0}{2.8} & \heatcellpm{47.5}{4.8} & \heatcellpm{52.0}{4.2} & \heatcellpm{54.5}{3.6} \\
Qwen3-8B & \heatcellpm{49.4}{2.1} & \heatcellpm{19.0}{3.6} & \heatcellpm{37.5}{1.7} & \heatcellpm{78.5}{1.7} & \heatcellpm{54.5}{5.2} & \heatcellpm{57.5}{1.7} \\
Devstral 2 123B & \heatcellpm{42.1}{2.1} & \heatcellpm{33.5}{1.7} & \heatcellpm{40.5}{9.5} & \heatcellpm{51.5}{1.7} & \heatcellpm{44.5}{3.8} & \heatcellpm{40.5}{4.6} \\
Ministral 3 8B & \heatcellpm{41.7}{0.9} & \heatcellpm{32.5}{3.0} & \heatcellpm{39.0}{1.0} & \heatcellpm{58.5}{1.7} & \heatcellpm{32.5}{3.8} & \heatcellpm{46.0}{4.2} \\
Qwen3-Coder-30B & \heatcellpm{38.7}{2.3} & \heatcellpm{27.5}{3.0} & \heatcellpm{30.0}{2.4} & \heatcellpm{53.0}{3.0} & \heatcellpm{46.0}{5.8} & \heatcellpm{37.0}{6.1} \\
\end{tabular}%
}
\end{table}

\begin{table}[H]
\centering
\scriptsize
\vspace{-1em}
\renewcommand{\arraystretch}{0.60}
\caption{Baseline per-property accuracy (\%) on our benchmark restricted to programs of 101--400 lines. Compared to Table~\ref{tab:baseline-heat}, violation detection collapses for most models while holds accuracy remains high.}
\label{tab:baseline-heat-long}
\resizebox{\columnwidth}{!}{%
\begin{tabular}{l|c|c|c|c|c|c}
\multicolumn{7}{c}{\textit{Property Holds (101--400 LOC)}} \\
\textbf{Model} & \textbf{Total} & \textbf{Mem. Safety} & \textbf{Overflow} & \textbf{Termin.} & \textbf{Reach.} & \textbf{Data Race} \\
\hline
Devstral 2 123B & \heatcellpm{99.0}{1.0} & \heatcell{100.0} & \heatcellpm{95.0}{5.0} & \heatcell{100.0} & \heatcell{100.0} & \heatcell{100.0} \\
Mistral Large 3 675B & \heatcellpm{98.5}{0.9} & \heatcell{100.0} & \heatcellpm{92.5}{4.3} & \heatcell{100.0} & \heatcell{100.0} & \heatcell{100.0} \\
Claude Opus 4.7 & \heatcell{98.0} & \heatcell{100.0} & \heatcell{90.0} & \heatcell{100.0} & \heatcell{100.0} & \heatcell{100.0} \\
DeepSeek V3.2 & \heatcellpm{97.4}{0.9} & \heatcell{100.0} & \heatcell{90.0} & \heatcellpm{97.2}{4.8} & \heatcell{100.0} & \heatcell{100.0} \\
Kimi K2.5 & \heatcellpm{96.9}{1.0} & \heatcell{100.0} & \heatcell{90.0} & \heatcellpm{94.4}{5.6} & \heatcell{100.0} & \heatcell{100.0} \\
Qwen3-Coder-30B & \heatcellpm{96.4}{3.0} & \heatcell{100.0} & \heatcellpm{97.5}{4.3} & \heatcell{100.0} & \heatcellpm{97.5}{4.3} & \heatcellpm{87.5}{8.3} \\
Ministral 3 8B & \heatcellpm{96.4}{0.9} & \heatcellpm{90.0}{7.1} & \heatcell{100.0} & \heatcell{100.0} & \heatcellpm{95.0}{5.0} & \heatcellpm{97.5}{4.3} \\
Qwen3-32B & \heatcellpm{97.4}{1.7} & \heatcell{100.0} & \heatcellpm{95.0}{5.0} & \heatcellpm{94.4}{5.6} & \heatcell{100.0} & \heatcellpm{97.5}{4.3} \\
DeepSeek R1 & \heatcellpm{94.4}{1.7} & \heatcellpm{97.5}{4.3} & \heatcellpm{87.5}{4.3} & \heatcellpm{97.2}{4.8} & \heatcellpm{90.0}{7.1} & \heatcell{100.0} \\
Claude Sonnet 4.5 & \heatcellpm{92.3}{1.7} & \heatcellpm{97.5}{4.3} & \heatcell{90.0} & \heatcellpm{91.7}{4.8} & \heatcell{100.0} & \heatcellpm{82.5}{4.3} \\
Qwen3-Coder-480B & \heatcellpm{90.8}{3.4} & \heatcellpm{77.5}{4.3} & \heatcell{90.0} & \heatcellpm{97.2}{4.8} & \heatcellpm{92.5}{8.3} & \heatcellpm{97.5}{4.3} \\
Qwen3-8B & \heatcellpm{89.3}{2.7} & \heatcellpm{97.5}{4.3} & \heatcellpm{90.0}{7.1} & \heatcell{100.0} & \heatcellpm{87.5}{4.3} & \heatcellpm{72.5}{8.3} \\
GPT-OSS 120B & \heatcellpm{88.8}{1.0} & \heatcell{100.0} & \heatcellpm{85.0}{5.0} & \heatcellpm{58.3}{4.8} & \heatcellpm{97.5}{4.3} & \heatcell{100.0} \\
GPT-OSS 20B & \heatcellpm{73.5}{3.2} & \heatcellpm{85.0}{5.0} & \heatcellpm{72.5}{10.9} & \heatcellpm{66.7}{11.1} & \heatcellpm{45.0}{15.0} & \heatcellpm{97.5}{4.3} \\
\end{tabular}%
}
\end{table}

\begin{table}[H]
\centering
\scriptsize
\renewcommand{\arraystretch}{0.60}
\resizebox{\columnwidth}{!}{%
\begin{tabular}{l|c|c|c|c|c|c}
\multicolumn{7}{c}{\textit{Property Violated (101--400 LOC)}} \\
\textbf{Model} & \textbf{Total} & \textbf{Mem. Safety} & \textbf{Overflow} & \textbf{Termin.} & \textbf{Reach.} & \textbf{Data Race} \\
\hline
GPT-OSS 120B & \heatcellpm{84.9}{0.9} & \heatcellpm{85.0}{5.0} & \heatcellpm{77.5}{4.3} & \heatcell{100.0} & \heatcell{70.0} & \heatcellpm{93.8}{10.8} \\
Claude Opus 4.7 & \heatcellpm{82.8}{1.7} & \heatcellpm{72.5}{4.3} & \heatcell{100.0} & \heatcellpm{70.0}{7.1} & \heatcell{80.0} & \heatcellpm{93.8}{6.2} \\
DeepSeek V3.2 & \heatcellpm{75.5}{7.1} & \heatcellpm{62.5}{19.2} & \heatcellpm{60.0}{12.2} & \heatcellpm{80.0}{10.0} & \heatcellpm{82.5}{10.9} & \heatcellpm{96.9}{5.4} \\
DeepSeek R1 & \heatcellpm{71.4}{3.7} & \heatcellpm{72.5}{13.0} & \heatcellpm{37.5}{4.3} & \heatcellpm{92.5}{8.3} & \heatcellpm{67.5}{8.3} & \heatcellpm{90.6}{10.4} \\
Kimi K2.5 & \heatcellpm{68.8}{2.9} & \heatcellpm{70.0}{7.1} & \heatcellpm{50.0}{7.1} & \heatcellpm{85.0}{5.0} & \heatcellpm{62.5}{10.9} & \heatcellpm{78.1}{5.4} \\
Claude Sonnet 4.5 & \heatcellpm{67.7}{3.8} & \heatcellpm{75.0}{5.0} & \heatcellpm{52.5}{4.3} & \heatcellpm{75.0}{5.0} & \heatcellpm{42.5}{10.9} & \heatcell{100.0} \\
GPT-OSS 20B & \heatcellpm{66.7}{4.4} & \heatcellpm{55.0}{8.7} & \heatcellpm{52.5}{10.9} & \heatcell{100.0} & \heatcellpm{52.5}{13.0} & \heatcellpm{75.0}{12.5} \\
Qwen3-Coder-480B & \heatcellpm{58.3}{5.1} & \heatcellpm{67.5}{4.3} & \heatcellpm{37.5}{4.3} & \heatcellpm{72.5}{10.9} & \heatcellpm{52.5}{13.0} & \heatcellpm{62.5}{8.8} \\
Qwen3-8B & \heatcellpm{34.9}{4.3} & \heatcellpm{12.5}{4.3} & \heatcellpm{7.5}{4.3} & \heatcellpm{82.5}{4.3} & \heatcellpm{42.5}{13.0} & \heatcellpm{28.1}{13.6} \\
Qwen3-32B & \heatcellpm{34.4}{1.8} & \heatcellpm{5.0}{5.0} & \heatcellpm{10.0}{7.1} & \heatcellpm{92.5}{8.3} & \heatcellpm{45.0}{5.0} & \heatcellpm{15.6}{5.4} \\
Ministral 3 8B & \heatcellpm{17.7}{4.3} & \heatcell{20.0} & \heatcellpm{15.0}{5.0} & \heatcellpm{27.5}{4.3} & \heatcellpm{15.0}{8.7} & \heatcellpm{9.4}{10.4} \\
Qwen3-Coder-30B & \heatcellpm{16.1}{4.0} & \heatcellpm{12.5}{4.3} & \heatcellpm{7.5}{4.3} & \heatcellpm{35.0}{5.0} & \heatcellpm{22.5}{14.8} & \heatcell{0.0} \\
Devstral 2 123B & \heatcellpm{13.5}{1.0} & \heatcellpm{12.5}{4.3} & \heatcellpm{5.0}{5.0} & \heatcellpm{20.0}{7.1} & \heatcellpm{22.5}{8.3} & \heatcellpm{6.2}{10.8} \\
Mistral Large 3 675B & \heatcellpm{12.5}{2.6} & \heatcellpm{17.5}{10.9} & \heatcell{10.0} & \heatcellpm{5.0}{5.0} & \heatcellpm{17.5}{4.3} & \heatcell{12.5} \\
\end{tabular}%
}
\end{table}

\subsection{Failure Analysis}
\label{app:failure-example}

We analyze failures from a single representative run (run\_1) across all 14 models, observing 1217 incorrect answers and 57 unparseable responses out of 6994 total evaluations. Among the incorrect answers, 1069 are false negatives (model says OKAY on a violated property) and 148 are false positives (model says ERROR on a property that holds), a 7.2:1 ratio confirming the systematic overoptimism discussed in the main text. To systematically categorize failure modes, we combine manual review of a subset of responses (representative examples below) with an LLM-as-a-judge approach for the full set: reviewing all 1217 incorrect responses manually is infeasible, so we use Claude Opus 4.7 to classify each parseable incorrect response into predefined categories. Tables~\ref{tab:fn-breakdown} and~\ref{tab:fp-breakdown} summarize the results for the 1069 parseable false negatives and 148 parseable false positives.

\paragraph{False Negatives.} The most common failure mode is \emph{missing edge cases} (44\%): models analyze the code but fail to consider the specific inputs or execution paths that trigger the violation. This is the dominant mode for memory safety (68\%) and data race (51\%) violations, where models typically trace the ``happy path'' without exploring adversarial inputs. The second most common is \emph{identified but dismissed} (27\%): models correctly identify the bug mechanism but rationalize it away. This is especially prevalent for memory safety (28\%) and data race (35\%), where models often appeal to practical runtime behavior rather than applying the formal specification. \emph{Incorrect reasoning} (19\%) is most common for reachability (33\%) and overflow (27\%), where models make control flow or arithmetic errors. \emph{Universal/existential confusion} (8\%) is concentrated almost entirely in termination (49\%), where models confuse ``can terminate'' with ``must terminate'' --- a qualitatively distinct failure reflecting a misunderstanding of the verification task rather than the code.

\begin{table}[H]
\centering
\scriptsize
\renewcommand{\arraystretch}{0.85}
\setlength{\tabcolsep}{4pt}
\caption{False negative failure modes classified by Claude Opus 4.7 as judge (all 14 models).}
\label{tab:fn-breakdown}
\resizebox{\columnwidth}{!}{%
\begin{tabular}{l@{\hspace{8pt}}rrrrr@{\hspace{6pt}}r}
\toprule
& \textbf{Missed edge case} & \textbf{Ident.\ but dismissed} & \textbf{Incorrect reasoning} & \textbf{$\forall$/$\exists$ confusion} & \textbf{Superficial} & \textbf{Total} \\
\midrule
Mem.\ Safety & 186 & 77 & 8 & 1 & 1 & 273 \\
Overflow & 85 & 68 & 57 & 4 & 0 & 214 \\
Termination & 9 & 25 & 36 & 66 & 0 & 136 \\
Reachability & 88 & 50 & 79 & 13 & 11 & 241 \\
Data Race & 106 & 72 & 27 & 0 & 0 & 205 \\
\midrule
\textbf{Total} & \textbf{474 (44\%)} & \textbf{292 (27\%)} & \textbf{207 (19\%)} & \textbf{84 (8\%)} & \textbf{12 (1\%)} & \textbf{1069} \\
\bottomrule
\end{tabular}%
}
\end{table}

\paragraph{False Positives.} The dominant failure mode is \emph{incorrect reasoning} (49\%): models make errors in their analysis that lead them to flag non-existent bugs. \emph{Overly conservative} judgments (26\%) are concentrated in overflow and data race, where models flag operations that are technically safe. \emph{Misunderstood specification} (19\%) is most common for termination, where models misinterpret the property being checked. \emph{Fabricated bugs} (5\%) are most common for data race, where models claim a race condition exists in code that is actually safe.

\begin{table}[H]
\centering
\scriptsize
\renewcommand{\arraystretch}{0.85}
\setlength{\tabcolsep}{4pt}
\caption{False positive failure modes classified by Claude Opus 4.7 as judge (all 14 models).}
\label{tab:fp-breakdown}
\resizebox{\columnwidth}{!}{%
\begin{tabular}{l@{\hspace{8pt}}rrrrr@{\hspace{6pt}}r}
\toprule
& \textbf{Incorrect reasoning} & \textbf{Misunderstood spec} & \textbf{Fabricated bug} & \textbf{Overly conservative} & \textbf{Superficial} & \textbf{Total} \\
\midrule
Mem.\ Safety & 8 & 5 & 2 & 5 & 0 & 20 \\
Overflow & 25 & 0 & 0 & 20 & 0 & 45 \\
Termination & 7 & 23 & 0 & 1 & 0 & 31 \\
Reachability & 21 & 0 & 0 & 4 & 1 & 26 \\
Data Race & 12 & 0 & 5 & 9 & 0 & 26 \\
\midrule
\textbf{Total} & \textbf{73 (49\%)} & \textbf{28 (19\%)} & \textbf{7 (5\%)} & \textbf{39 (26\%)} & \textbf{1 (1\%)} & \textbf{148} \\
\bottomrule
\end{tabular}%
}
\end{table}

The following examples illustrate these failure modes in detail.

\subsubsection{Termination: Unsigned Wrapping}

The model correctly identifies unsigned integer wrapping but incorrectly reasons that the wrapped sequence must eventually reach zero, missing a non-termination bug. This response is from Mistral Large 3 675B on \texttt{termination-crafted/Cairo\_step2-3.c}.

\textbf{Program:}
\begin{lstlisting}[language=C, basicstyle=\small\ttfamily, breaklines=true]
typedef enum {false, true} bool;

extern unsigned int __VERIFIER_nondet_uint(void);

int main()
{
    unsigned int x;
    x = __VERIFIER_nondet_uint();
    if (x > 0) {
        while (x != 0) {
            x = x - 2;
        }
    }
    return 0;
}
\end{lstlisting}

\textbf{Model response (excerpt):}
\begin{quote}
\itshape
If \texttt{x} is odd, it will eventually become 1, and then the next subtraction will underflow (since \texttt{unsigned int} wraps around). For example: If \texttt{x = 1}, then \texttt{x = x - 2} will wrap around to \texttt{UINT\_MAX - 1} [...] However, this is not a problem for termination because [...] repeated subtractions of 2 will eventually bring it to 0 (since \texttt{UINT\_MAX} is odd, and subtracting 2 repeatedly from an odd number will eventually reach 0).

Final answer: OKAY
\end{quote}

\textbf{Ground truth:} Property violated (non-termination). When \texttt{x} is odd, subtracting 2 yields $1 \to 2^{32}-1 \to 2^{32}-3 \to \ldots$, cycling through odd values and never reaching 0. The model correctly identifies the unsigned wrapping but incorrectly claims the sequence must reach 0. In fact, since $2^{32}-1$ is odd and subtracting 2 preserves parity, the value cycles through odd numbers indefinitely.

\subsubsection{Reachability: Missing Fall-Through}

The model fails to trace control flow after a recursive call returns, missing that execution continues to an error label rather than returning from the function. This response is from Ministral 3 8B on \texttt{recursive-simple/afterrec-1.c}.

\textbf{Program:}
\begin{lstlisting}[language=C, basicstyle=\small\ttfamily, breaklines=true]
void reach_error() {
  __assert_fail("0", "afterrec-1.c", 3, "reach_error");
}

void f(int n) {
  if (n<3) return;
  n--;
  f(n);
  ERROR: {reach_error();abort();}
}

int main(void) {
  f(4);
}
\end{lstlisting}

\textbf{Model response (excerpt):}
\begin{quote}
\itshape
The recursion terminates naturally when \texttt{n} becomes less than 3. The \texttt{reach\_error} function is only called if the code reaches the \texttt{ERROR} label, which it never does in this execution flow.

Final answer: OKAY
\end{quote}

\textbf{Ground truth:} Property violated (\texttt{reach\_error} is called). When \texttt{f(3)} calls \texttt{f(2)}, the inner call returns immediately (since $2 < 3$), but execution in \texttt{f(3)} continues past the recursive call and falls through to the \texttt{ERROR} label. The model incorrectly assumes the function returns after the recursive call.

\subsubsection{Memory Safety: Out-of-Bounds Write}

The model correctly identifies that a \texttt{memset} call writes one byte beyond the struct, but rationalizes the violation away by arguing it is ``still within the stack.'' This response is from Devstral 2 123B on \texttt{ldv-memsafety/memset2\_-write.c}.

\textbf{Program:}
\begin{lstlisting}[language=C, basicstyle=\small\ttfamily, breaklines=true]
typedef unsigned int __u32;
struct compstat {
   __u32 unc_bytes; __u32 unc_packets;
   __u32 comp_bytes; __u32 comp_packets;
   __u32 inc_bytes; __u32 inc_packets;
   __u32 in_count; __u32 bytes_out;
   double ratio;
};

int main() {
  struct compstat cstats;
  memset((void *)(&cstats), 0, 41UL);
  return 0;
}
\end{lstlisting}

\textbf{Model response (excerpt):}
\begin{quote}
\itshape
The size 41 is slightly larger than the actual size of the struct (40 bytes), but since \texttt{cstats} is a local variable and not a dynamically allocated block, accessing one byte beyond the allocated memory (which is still within the stack) does not cause a segmentation fault. [...] This is acceptable.

Final answer: OKAY
\end{quote}

\textbf{Ground truth:} Property violated (invalid dereference). The struct is 40 bytes, but \texttt{memset} writes 41 bytes, overwriting one byte beyond the struct's bounds. The model correctly computes the size mismatch but incorrectly dismisses it as harmless because the memory is on the stack. While the write may not cause a crash in practice (the extra byte likely lands on accessible stack memory), the SV-COMP memory safety property requires that every dereference stays within the bounds of the allocated object. The model defaults to reasoning about practical runtime behavior rather than applying the formal specification it was asked to check.

\subsubsection{Termination: Oscillating Variables}

Multiple models fail to detect non-termination when two variables swap values in a way that prevents monotonic decrease. 7 of 14 models answer incorrectly, including Claude Sonnet 4.5 and Mistral Large 3 675B. Claude Opus 4.7 and DeepSeek R1 correctly identify the issue. Responses shown are from \texttt{termination-restricted-15/Flip2.c}.

\textbf{Program:}
\begin{lstlisting}[language=C, basicstyle=\footnotesize\ttfamily, breaklines=true]
typedef enum {false,true} bool;
extern int __VERIFIER_nondet_int(void);

int main() {
    int i, j, t;
    i = __VERIFIER_nondet_int();
    j = __VERIFIER_nondet_int();
    t = 0;
    while (i > 0 && j > 0) {
        if (i < j) { t = i; i = j; j = t; }
        else if (i > j) { j = i; }
        else { i = i-1; }
    }
    return 0;
}
\end{lstlisting}

\textbf{Claude Sonnet 4.5 response (excerpt):}
\begin{quote}
\itshape
The key insight: Eventually, through this process, \texttt{i} keeps getting decremented (at least every few iterations), and once \texttt{i} $\leq$ 0, the loop terminates. Since \texttt{i} is decremented regularly and will eventually become $\leq$ 0, the loop must terminate.

Final answer: OKAY
\end{quote}

\textbf{DeepSeek R1 response (excerpt):}
\begin{quote}
\itshape
[\ldots] the loop runs indefinitely. This occurs when \texttt{i} and \texttt{j} alternate between values like 4 and 5, leading to no net decrease in either variable.

Final answer: ERROR
\end{quote}

\textbf{Ground truth:} Property violated (non-termination). When \texttt{i < j}, the swap makes \texttt{i} the larger value; then \texttt{i > j} triggers \texttt{j = i}, equalizing them; then \texttt{i == j} decrements \texttt{i} by 1, making \texttt{i < j} again. For example, starting with \texttt{i=4, j=5}: the cycle \texttt{(4,5) $\to$ (5,4) $\to$ (5,5) $\to$ (4,5)} repeats indefinitely. Claude Sonnet 4.5 incorrectly assumes \texttt{i} decreases monotonically.

\subsubsection{Termination: Nondeterministic Loop Exit}

Mistral Large 3 675B and several other models fail to recognize that a nondeterministic loop condition can prevent termination. This response is from \texttt{loop-lit/mine2017-ex4.7.c}.

\textbf{Program:}
\begin{lstlisting}[language=C, basicstyle=\footnotesize\ttfamily, breaklines=true]
extern _Bool __VERIFIER_nondet_bool();
extern void abort(void);
void reach_error() { assert(0); }
void __VERIFIER_assert(int cond) {
  if(!(cond)) { ERROR: {reach_error();abort();} } }

int main() {
  int x = 0;
  while (__VERIFIER_nondet_bool() == 0) {
    __VERIFIER_assert(0 <= x);
    __VERIFIER_assert(x <= 40);
    if (__VERIFIER_nondet_bool() == 0) {
      x++;
      if (x > 40) x = 0;
    }
  }
  return 0;
}
\end{lstlisting}

\textbf{Mistral Large 3 675B response (excerpt):}
\begin{quote}
\itshape
The loop will terminate because \texttt{\_\_VERIFIER\_nondet\_bool()} can return \texttt{1} (terminating the loop) at any point.

Final answer: OKAY
\end{quote}

\textbf{Ground truth:} Property violated (non-termination). The loop condition depends on \texttt{\_\_VERIFIER\_nondet\_bool()}, which returns an arbitrary boolean. There exists an execution path where it always returns \texttt{0}, causing the loop to run forever. The model confuses ``can terminate'' with ``must terminate'': the property requires that \emph{every} execution path reaches the end of the program, not just that some path does.

\section{Training Details and Experiment Configurations}
\label{app:training-details}

\paragraph{Inference.} Evaluation uses vLLM with tensor parallelism across 8 GPUs, greedy decoding (temperature 0), and a maximum output length of 8192 tokens. Prompts exceeding 4096 tokens are excluded.

\subsection{Experiment Configurations}
\label{app:experiment-configs}

Table~\ref{tab:experiment-configs} lists all training configurations evaluated in this work. Unless otherwise noted, CPT configurations use Qwen3-8B, AdamW, LR $1{\times}10^{-5}$, cosine schedule, 1 epoch, batch size 64, FSDP across 8 NVIDIA A100 GPUs (40GB for 8B training and evaluation, 80GB for 32B evaluation). We use a single epoch without a validation set; the learning rate was chosen based on a single comparison (LR $5{\times}10^{-6}$ underperformed on the best configuration) rather than a systematic search. ``Thinking'' refers to Qwen3's internal chain-of-thought mode; ``no thinking'' disables it. All evaluations use 4 runs; we report mean$\pm$std.

\begin{table}[H]
\centering
\scriptsize
\renewcommand{\arraystretch}{0.72}
\caption{Training configurations and evaluation settings.}
\label{tab:experiment-configs}
\resizebox{\columnwidth}{!}{%
\begin{tabular}{l|l}
\textbf{ID} & \textbf{Description} \\
\hline
\multicolumn{2}{c}{\textit{Baselines}} \\
\hline
\texttt{qwen3-8B-base-no-thinking} & Qwen3-8B base model (no fine-tuning). No thinking. \\
\texttt{qwen3-8B-base-thinking} & Qwen3-8B base model. Thinking enabled. 14k max tokens. \\
\texttt{qwen3-32b-base-no-thinking} & Qwen3-32B base model (no fine-tuning). No thinking. \\
\texttt{qwen3-32b-base-thinking} & Qwen3-32B base model. Thinking enabled. \\
\hline
\multicolumn{2}{c}{\textit{CPT on HTML Traces}} \\
\hline
\texttt{qwen3-8b-cpt-html-bugs-no-thinking} & Full CPT on 3,208 HTML bug traces (manifest violations only). No thinking. \\
\texttt{qwen3-8b-cpt-html-bugs-thinking} & Same model. Thinking enabled. \\
\texttt{qwen3-8b-cpt-html-bugs-lr5e6-no-thinking} & Same as html-bugs but LR $5{\times}10^{-6}$. No thinking. \\
\texttt{qwen3-8b-cpt-html-expanded-no-thinking} & CPT on 5,048 HTML traces (754 manifest + 4,294 latent; manifest subset sampled to control the manifest/latent ratio). No thinking. \\
\texttt{qwen3-8b-cpt-html-expanded-thinking} & Same model. Thinking enabled. \\
\texttt{qwen3-8b-cpt-html-all-no-thinking} & CPT on all 34,495 HTML traces (bugs + non-bugs). No thinking. \\
\texttt{qwen3-8b-cpt-html-all-thinking} & Same model. Thinking enabled. \\
\texttt{qwen3-8b-cpt-html-mixed-thinking} & CPT on 1,754 HTML traces (754 manifest + 1k latent). Thinking. \\
\hline
\multicolumn{2}{c}{\textit{CPT on Plain-Text Traces}} \\
\hline
\texttt{qwen3-8B-soteria-bug-traces-plaintext-no-thinking} & CPT on 3,208 plain-text bug traces (HTML removed, manifest only). No thinking. \\
\texttt{qwen3-8b-plaintext-bugs-thinking} & Same model. Thinking. Collapsed: 45\% unparsed. \\
\texttt{qwen3-8b-cpt-bugs-only-no-thinking} & CPT on 5,048 plain-text traces (754 manifest + 4,294 latent). No thinking. \\
\texttt{qwen3-8b-cpt-bugs-only-3ep-no-thinking} & Same dataset, 3 epochs. No thinking. \\
\texttt{qwen3-8b-cpt-filtered-no-thinking} & CPT on 16,234 plain-text traces (filtered, no incomplete). No thinking. \\
\texttt{qwen3-8B-soteria-all-traces-plaintext-no-thinking} & CPT on 34,495 plain-text traces (all). No thinking. \\
\texttt{qwen3-8B-soteria-balanced-no-verdict-no-thinking} & CPT on 6,416 plain-text traces (50/50, verdict removed). No thinking. \\
\texttt{qwen3-8b-cpt-bugs-stripped-no-thinking} & CPT on 5,048 stripped traces (error-relevant lines only, 72\% reduction). No thinking. \\
\hline
\multicolumn{2}{c}{\textit{CPT on Other Data}} \\
\hline
\texttt{qwen3-8b-cpt-code-only-no-thinking} & CPT on 3,208 C source files only (no traces). Ablation. No thinking. \\
\texttt{qwen3-8b-cpt-informalized-no-thinking} & CPT on 5,048 traces informalized by Claude Opus 4.7. Collapsed: 62\% unparsed. \\
\hline
\multicolumn{2}{c}{\textit{CPT + SFT (Two-Stage)}} \\
\hline
\texttt{qwen3-8b-cpt-sft-verdicts-no-thinking} & CPT on 5,048 bug traces, then LoRA SFT r4 on 1,508 verdict examples (input: C code + property question; output: brief reasoning + ``ERROR''/``OKAY'' verdict). No thinking. \\
\hline
\multicolumn{2}{c}{\textit{LoRA SFT (No CPT)}} \\
\hline
\texttt{qwen3-8b-lora-sft-v2-no-thinking} & LoRA SFT on 13,155 examples (5.7\% bugs). Input: C code + ``trace the execution'' instruction; output: full Soteria trace. r16, $\alpha$32, LR $2{\times}10^{-4}$. No thinking. \\
\texttt{qwen3-8b-lora-sft-v3-no-thinking} & LoRA SFT r4, $\alpha$8 on 2,504 examples (30\% bugs). LR $1{\times}10^{-4}$. No thinking. \\
\texttt{qwen3-8b-lora-sft-v4-no-thinking} & LoRA SFT r4, $\alpha$8 on 5,004 examples (15\% bugs). LR $1{\times}10^{-4}$. No thinking. \\
\texttt{qwen3-8b-lora-sft-v5-no-thinking} & LoRA SFT r8, $\alpha$16 on 5,004 examples. LR $1{\times}10^{-4}$. Collapsed: 70\% unparsed. \\
\texttt{qwen3-8b-lora-sft-v6-no-thinking} & LoRA SFT r8, $\alpha$16 on 5,004 examples. LR $2{\times}10^{-5}$. No thinking. \\
\hline
\multicolumn{2}{c}{\textit{Prompting Baselines (No Training)}} \\
\hline
\texttt{qwen3-8B-base-fewshot-no-thinking} & Base model with 2 Soteria trace examples in prompt (1 bug, 1 no-bug). No thinking. \\
\end{tabular}%
}
\end{table}

\subsection{Detailed Training Results}
\label{app:training-results-detailed}

Tables~\ref{tab:training-heat-holds} and~\ref{tab:training-heat-violated} show per-property accuracy for all training configurations on our benchmark (1--400 LOC). Each value is the mean across 4 runs; $\pm$ indicates standard deviation.

\begin{table}[H]
\centering
\scriptsize
\renewcommand{\arraystretch}{0.65}
\caption{Per-property accuracy (\%) on our benchmark --- property holds (1--400 LOC).}
\label{tab:training-heat-holds}
\resizebox{\columnwidth}{!}{%
\begin{tabular}{l|c|c|c|c|c|c}
\textbf{Configuration} & \textbf{Total} & \textbf{Mem.} & \textbf{Over.} & \textbf{Term.} & \textbf{Reach.} & \textbf{Race} \\
\hline
\texttt{qwen3-32b-base-thinking} & \heatcellpm{95.2}{0.3} & \heatcell{98.0} & \heatcellpm{93.0}{1.7} & \heatcellpm{92.5}{2.2} & \heatcellpm{93.5}{0.9} & \heatcellpm{99.0}{1.0} \\
\texttt{qwen3-32b-base-no-thinking} & \heatcellpm{95.4}{0.6} & \heatcellpm{97.5}{0.9} & \heatcellpm{94.0}{3.7} & \heatcellpm{94.5}{0.9} & \heatcellpm{94.0}{2.4} & \heatcellpm{97.0}{1.0} \\
\texttt{qwen3-8b-lora-sft-v4-no-thinking} & \heatcellpm{92.0}{0.8} & \heatcell{98.0} & \heatcellpm{87.0}{3.6} & \heatcellpm{97.5}{0.9} & \heatcellpm{93.5}{0.9} & \heatcellpm{84.0}{3.2} \\
\texttt{qwen3-8b-lora-sft-v3-no-thinking} & \heatcellpm{91.7}{1.5} & \heatcellpm{96.5}{1.7} & \heatcellpm{89.0}{1.7} & \heatcellpm{97.0}{3.0} & \heatcellpm{94.0}{1.4} & \heatcellpm{82.0}{2.4} \\
\texttt{qwen3-8B-base-no-thinking} & \heatcellpm{90.7}{0.6} & \heatcellpm{96.0}{1.4} & \heatcellpm{89.0}{2.2} & \heatcellpm{96.0}{1.4} & \heatcellpm{89.5}{4.6} & \heatcellpm{83.0}{5.0} \\
\texttt{qwen3-8B-base-fewshot-no-thinking} & \heatcellpm{92.2}{0.9} & \heatcellpm{88.5}{4.6} & \heatcellpm{95.0}{4.1} & \heatcellpm{96.0}{1.4} & \heatcellpm{93.0}{2.2} & \heatcellpm{88.5}{2.6} \\
\texttt{qwen3-8b-lora-sft-v6-no-thinking} & \heatcellpm{89.8}{1.8} & \heatcellpm{95.5}{1.7} & \heatcellpm{93.5}{2.2} & \heatcellpm{96.0}{1.4} & \heatcellpm{86.0}{4.5} & \heatcellpm{78.0}{2.0} \\
\texttt{qwen3-8b-cpt-code-only-no-thinking} & \heatcellpm{87.4}{0.4} & \heatcellpm{94.0}{2.4} & \heatcellpm{91.0}{3.3} & \heatcellpm{96.5}{0.9} & \heatcellpm{83.5}{3.3} & \heatcellpm{72.0}{2.4} \\
\texttt{qwen3-8B-base-thinking} & \heatcellpm{85.4}{1.7} & \heatcellpm{95.5}{1.7} & \heatcellpm{79.0}{3.3} & \heatcellpm{94.5}{1.7} & \heatcellpm{81.5}{1.7} & \heatcellpm{76.5}{3.8} \\
\texttt{qwen3-8b-cpt-bugs-stripped-no-thinking} & \heatcellpm{86.4}{1.2} & \heatcellpm{95.5}{1.7} & \heatcellpm{92.0}{3.7} & \heatcellpm{95.5}{1.7} & \heatcellpm{82.0}{2.4} & \heatcellpm{67.0}{4.1} \\
\texttt{qwen3-8b-cpt-sft-verdicts-no-thinking} & \heatcellpm{86.8}{1.0} & \heatcellpm{92.0}{2.8} & \heatcellpm{86.0}{2.8} & \heatcellpm{95.0}{1.7} & \heatcellpm{91.5}{2.2} & \heatcellpm{69.5}{3.8} \\
\texttt{qwen3-8b-cpt-html-all-no-thinking} & \heatcellpm{83.8}{1.8} & \heatcellpm{83.0}{4.1} & \heatcellpm{88.5}{1.7} & \heatcellpm{94.5}{2.6} & \heatcellpm{83.0}{1.0} & \heatcellpm{70.0}{7.5} \\
\texttt{qwen3-8b-cpt-html-expanded-no-thinking} & \heatcellpm{83.7}{1.5} & \heatcellpm{93.5}{1.7} & \heatcellpm{86.5}{2.2} & \heatcellpm{91.5}{0.9} & \heatcellpm{79.0}{3.6} & \heatcellpm{68.0}{4.7} \\
\texttt{qwen3-8b-cpt-bugs-only-no-thinking} & \heatcellpm{84.6}{1.8} & \heatcellpm{91.5}{3.0} & \heatcellpm{89.0}{2.2} & \heatcellpm{96.0}{1.4} & \heatcellpm{88.0}{3.2} & \heatcellpm{58.5}{5.4} \\
\texttt{qwen3-8b-cpt-html-expanded-thinking} & \heatcellpm{78.7}{1.2} & \heatcellpm{84.5}{4.3} & \heatcellpm{67.0}{1.0} & \heatcellpm{87.5}{1.7} & \heatcellpm{80.5}{4.6} & \heatcellpm{74.0}{4.2} \\
\texttt{qwen3-8b-cpt-html-bugs-thinking} & \heatcellpm{78.0}{2.1} & \heatcellpm{79.5}{3.0} & \heatcellpm{71.0}{4.1} & \heatcellpm{83.0}{2.2} & \heatcellpm{80.5}{3.6} & \heatcellpm{76.0}{5.1} \\
\texttt{qwen3-8b-cpt-html-all-thinking} & \heatcellpm{77.9}{1.4} & \heatcellpm{85.5}{5.5} & \heatcellpm{79.5}{2.6} & \heatcellpm{85.0}{4.6} & \heatcellpm{69.5}{3.0} & \heatcell{70.0} \\
\texttt{qwen3-8b-cpt-html-mixed-thinking} & \heatcellpm{77.2}{1.3} & \heatcellpm{80.0}{5.8} & \heatcellpm{66.0}{2.4} & \heatcellpm{89.5}{3.8} & \heatcellpm{80.0}{3.2} & \heatcellpm{70.5}{4.3} \\
\texttt{qwen3-8b-cpt-html-bugs-lr5e6-no-thinking} & \heatcellpm{79.9}{1.2} & \heatcellpm{93.5}{4.3} & \heatcellpm{75.0}{5.4} & \heatcellpm{93.5}{1.7} & \heatcellpm{79.5}{0.9} & \heatcellpm{58.0}{5.1} \\
\texttt{qwen3-8b-cpt-filtered-no-thinking} & \heatcellpm{82.5}{0.9} & \heatcellpm{93.5}{0.9} & \heatcellpm{83.0}{3.3} & \heatcellpm{92.0}{1.4} & \heatcellpm{83.0}{1.7} & \heatcellpm{61.0}{1.0} \\
\texttt{qwen3-8b-cpt-html-bugs-no-thinking} & \heatcellpm{80.0}{0.7} & \heatcellpm{93.0}{1.7} & \heatcellpm{79.0}{1.0} & \heatcellpm{93.5}{1.7} & \heatcellpm{77.0}{2.2} & \heatcellpm{57.5}{4.3} \\
\texttt{qwen3-8B-soteria-all-traces-plaintext-no-thinking} & \heatcellpm{80.9}{1.2} & \heatcellpm{85.0}{1.7} & \heatcellpm{92.5}{2.6} & \heatcellpm{91.5}{3.0} & \heatcellpm{86.5}{5.7} & \heatcellpm{49.0}{6.6} \\
\texttt{qwen3-8b-cpt-bugs-only-3ep-no-thinking} & \heatcellpm{79.5}{1.8} & \heatcellpm{95.5}{0.9} & \heatcellpm{81.0}{3.3} & \heatcellpm{89.0}{3.0} & \heatcellpm{78.5}{3.3} & \heatcellpm{53.5}{7.4} \\
\texttt{qwen3-8b-lora-sft-v2-no-thinking} & \heatcellpm{77.9}{1.0} & \heatcellpm{82.5}{0.9} & \heatcellpm{72.5}{4.1} & \heatcellpm{87.0}{3.0} & \heatcellpm{90.5}{3.0} & \heatcellpm{57.0}{1.7} \\
\texttt{qwen3-8B-soteria-balanced-no-verdict-no-thinking} & \heatcellpm{75.0}{1.5} & \heatcellpm{82.0}{3.2} & \heatcellpm{70.0}{3.7} & \heatcellpm{87.5}{1.7} & \heatcellpm{79.0}{6.7} & \heatcellpm{56.5}{3.6} \\
\texttt{qwen3-8B-soteria-bug-traces-plaintext-no-thinking} & \heatcellpm{71.6}{2.9} & \heatcellpm{81.5}{1.7} & \heatcellpm{60.0}{5.1} & \heatcellpm{88.0}{3.5} & \heatcellpm{76.0}{6.8} & \heatcellpm{52.5}{5.9} \\
\texttt{qwen3-8b-plaintext-bugs-thinking} & \heatcellpm{54.6}{2.1} & \heatcellpm{70.5}{3.8} & \heatcellpm{47.5}{5.4} & \heatcellpm{60.5}{6.7} & \heatcellpm{53.5}{0.9} & \heatcellpm{41.0}{3.0} \\
\texttt{qwen3-8b-cpt-informalized-no-thinking} & \heatcellpm{28.9}{1.0} & \heatcellpm{13.0}{1.7} & \heatcellpm{25.5}{4.6} & \heatcellpm{36.0}{4.7} & \heatcellpm{46.0}{2.0} & \heatcellpm{24.0}{3.7} \\
\texttt{qwen3-8b-lora-sft-v5-no-thinking} & \heatcellpm{26.3}{1.4} & \heatcellpm{35.0}{2.2} & \heatcellpm{30.5}{3.3} & \heatcellpm{30.5}{3.0} & \heatcellpm{17.5}{2.2} & \heatcellpm{18.0}{2.4} \\
\end{tabular}%
}
\end{table}

\begin{table}[H]
\centering
\scriptsize
\renewcommand{\arraystretch}{0.65}
\caption{Per-property accuracy (\%) on our benchmark --- property violated (1--400 LOC).}
\label{tab:training-heat-violated}
\resizebox{\columnwidth}{!}{%
\begin{tabular}{l|c|c|c|c|c|c}
\textbf{Configuration} & \textbf{Total} & \textbf{Mem.} & \textbf{Over.} & \textbf{Term.} & \textbf{Reach.} & \textbf{Race} \\
\hline
\texttt{qwen3-32b-base-thinking} & \heatcellpm{77.8}{0.7} & \heatcellpm{69.0}{3.0} & \heatcellpm{72.5}{1.7} & \heatcellpm{95.5}{1.7} & \heatcellpm{77.5}{0.9} & \heatcellpm{74.5}{1.7} \\
\texttt{qwen3-8b-cpt-html-bugs-thinking} & \heatcellpm{67.3}{1.6} & \heatcellpm{47.0}{1.7} & \heatcellpm{66.0}{2.4} & \heatcellpm{94.5}{1.7} & \heatcellpm{60.0}{1.4} & \heatcellpm{69.0}{8.5} \\
\texttt{qwen3-8b-cpt-html-expanded-thinking} & \heatcellpm{62.2}{0.9} & \heatcellpm{33.5}{1.7} & \heatcellpm{62.0}{3.2} & \heatcellpm{89.0}{1.0} & \heatcellpm{58.0}{5.7} & \heatcellpm{68.5}{3.0} \\
\texttt{qwen3-8b-cpt-html-mixed-thinking} & \heatcellpm{62.0}{1.2} & \heatcellpm{36.0}{5.5} & \heatcellpm{60.0}{2.4} & \heatcellpm{90.0}{3.2} & \heatcellpm{53.5}{4.1} & \heatcellpm{70.5}{4.3} \\
\texttt{qwen3-8b-cpt-html-bugs-no-thinking} & \heatcellpm{56.7}{1.0} & \heatcellpm{39.0}{4.1} & \heatcellpm{51.0}{3.0} & \heatcellpm{79.5}{1.7} & \heatcellpm{48.5}{3.0} & \heatcellpm{65.5}{3.0} \\
\texttt{qwen3-32b-base-no-thinking} & \heatcellpm{56.9}{1.5} & \heatcellpm{39.0}{4.4} & \heatcellpm{52.5}{3.6} & \heatcellpm{88.5}{1.7} & \heatcellpm{55.0}{4.1} & \heatcellpm{49.5}{2.2} \\
\texttt{qwen3-8b-cpt-html-bugs-lr5e6-no-thinking} & \heatcellpm{53.6}{1.7} & \heatcellpm{24.5}{3.6} & \heatcellpm{55.0}{6.4} & \heatcellpm{78.0}{2.8} & \heatcellpm{44.0}{3.7} & \heatcellpm{66.5}{4.8} \\
\texttt{qwen3-8B-soteria-bug-traces-plaintext-no-thinking} & \heatcellpm{53.7}{0.3} & \heatcellpm{22.0}{1.4} & \heatcellpm{68.0}{3.7} & \heatcellpm{66.0}{4.0} & \heatcellpm{39.5}{4.6} & \heatcellpm{73.0}{3.0} \\
\texttt{qwen3-8b-cpt-html-expanded-no-thinking} & \heatcellpm{50.2}{1.3} & \heatcellpm{17.5}{1.7} & \heatcellpm{51.5}{4.6} & \heatcellpm{73.5}{3.0} & \heatcellpm{44.0}{1.4} & \heatcellpm{64.5}{3.6} \\
\texttt{qwen3-8b-cpt-bugs-only-3ep-no-thinking} & \heatcellpm{51.2}{1.9} & \heatcellpm{18.0}{3.5} & \heatcellpm{48.0}{2.4} & \heatcellpm{70.0}{4.5} & \heatcellpm{44.5}{7.0} & \heatcellpm{75.5}{6.1} \\
\texttt{qwen3-8b-cpt-html-all-thinking} & \heatcellpm{48.4}{1.6} & \heatcellpm{26.5}{5.0} & \heatcellpm{35.5}{1.7} & \heatcellpm{85.0}{6.4} & \heatcellpm{46.0}{4.7} & \heatcellpm{49.0}{2.2} \\
\texttt{qwen3-8B-base-no-thinking} & \heatcellpm{49.4}{2.1} & \heatcellpm{19.0}{3.6} & \heatcellpm{37.5}{1.7} & \heatcellpm{78.5}{1.7} & \heatcellpm{54.5}{5.2} & \heatcellpm{57.5}{1.7} \\
\texttt{qwen3-8B-base-fewshot-no-thinking} & \heatcellpm{41.1}{1.0} & \heatcellpm{24.0}{2.0} & \heatcellpm{26.5}{1.7} & \heatcellpm{73.0}{1.7} & \heatcellpm{36.0}{3.2} & \heatcellpm{46.0}{3.2} \\
\texttt{qwen3-8b-cpt-bugs-only-no-thinking} & \heatcellpm{49.4}{1.5} & \heatcellpm{14.0}{2.8} & \heatcellpm{50.5}{5.0} & \heatcellpm{70.5}{2.6} & \heatcellpm{42.5}{4.8} & \heatcellpm{69.5}{3.8} \\
\texttt{qwen3-8B-base-thinking} & \heatcellpm{48.0}{2.0} & \heatcellpm{24.5}{2.2} & \heatcellpm{44.5}{4.6} & \heatcellpm{82.5}{3.3} & \heatcellpm{43.5}{3.0} & \heatcellpm{45.0}{4.1} \\
\texttt{qwen3-8b-lora-sft-v2-no-thinking} & \heatcellpm{47.6}{0.5} & \heatcellpm{46.5}{0.9} & \heatcellpm{50.0}{2.8} & \heatcellpm{63.0}{3.6} & \heatcellpm{24.5}{2.6} & \heatcellpm{54.0}{1.4} \\
\texttt{qwen3-8b-cpt-filtered-no-thinking} & \heatcellpm{47.5}{0.7} & \heatcellpm{16.0}{2.0} & \heatcellpm{39.5}{2.6} & \heatcellpm{69.0}{4.1} & \heatcellpm{43.0}{3.3} & \heatcellpm{70.0}{5.7} \\
\texttt{qwen3-8b-lora-sft-v6-no-thinking} & \heatcellpm{46.3}{1.6} & \heatcellpm{17.0}{1.0} & \heatcellpm{31.5}{7.4} & \heatcellpm{73.0}{4.1} & \heatcellpm{49.5}{2.2} & \heatcellpm{60.5}{2.2} \\
\texttt{qwen3-8b-cpt-bugs-stripped-no-thinking} & \heatcellpm{40.4}{1.5} & \heatcellpm{13.5}{2.2} & \heatcellpm{30.5}{2.6} & \heatcellpm{61.0}{3.0} & \heatcellpm{37.5}{7.4} & \heatcellpm{59.5}{1.7} \\
\texttt{qwen3-8B-soteria-balanced-no-verdict-no-thinking} & \heatcellpm{42.2}{1.6} & \heatcellpm{14.5}{1.7} & \heatcellpm{49.5}{1.7} & \heatcellpm{47.0}{3.0} & \heatcellpm{40.5}{4.1} & \heatcellpm{59.5}{3.8} \\
\texttt{qwen3-8b-cpt-code-only-no-thinking} & \heatcellpm{40.2}{2.7} & \heatcellpm{22.5}{3.8} & \heatcellpm{33.0}{3.3} & \heatcellpm{59.0}{7.3} & \heatcellpm{43.0}{2.2} & \heatcellpm{43.5}{1.7} \\
\texttt{qwen3-8B-soteria-all-traces-plaintext-no-thinking} & \heatcellpm{41.9}{1.2} & \heatcellpm{13.0}{2.2} & \heatcellpm{25.5}{4.3} & \heatcellpm{65.5}{3.8} & \heatcellpm{35.5}{2.6} & \heatcellpm{70.0}{5.8} \\
\texttt{qwen3-8b-cpt-html-all-no-thinking} & \heatcellpm{39.9}{2.3} & \heatcellpm{20.5}{4.8} & \heatcellpm{25.5}{4.1} & \heatcellpm{65.5}{3.0} & \heatcellpm{32.0}{3.7} & \heatcellpm{56.0}{4.9} \\
\texttt{qwen3-8b-lora-sft-v3-no-thinking} & \heatcellpm{41.2}{1.0} & \heatcellpm{12.5}{0.9} & \heatcellpm{29.0}{1.7} & \heatcellpm{62.0}{4.5} & \heatcellpm{44.5}{2.6} & \heatcellpm{58.0}{3.2} \\
\texttt{qwen3-8b-cpt-sft-verdicts-no-thinking} & \heatcellpm{38.0}{1.8} & \heatcellpm{8.5}{2.2} & \heatcellpm{35.0}{5.4} & \heatcellpm{50.0}{4.5} & \heatcellpm{37.5}{2.6} & \heatcellpm{59.0}{1.7} \\
\texttt{qwen3-8b-plaintext-bugs-thinking} & \heatcellpm{33.2}{3.5} & \heatcellpm{22.0}{4.5} & \heatcellpm{23.5}{4.6} & \heatcellpm{62.5}{4.6} & \heatcellpm{14.5}{4.8} & \heatcellpm{43.5}{4.3} \\
\texttt{qwen3-8b-lora-sft-v4-no-thinking} & \heatcellpm{33.4}{1.9} & \heatcell{22.0} & \heatcellpm{37.0}{4.1} & \heatcellpm{47.0}{3.3} & \heatcellpm{30.0}{2.4} & \heatcellpm{31.0}{1.0} \\
\texttt{qwen3-8b-cpt-informalized-no-thinking} & \heatcellpm{21.5}{0.9} & \heatcellpm{5.5}{0.9} & \heatcellpm{33.0}{1.0} & \heatcellpm{24.0}{2.0} & \heatcellpm{32.5}{1.7} & \heatcellpm{12.5}{3.0} \\
\texttt{qwen3-8b-lora-sft-v5-no-thinking} & \heatcellpm{5.5}{0.8} & \heatcellpm{7.0}{1.7} & \heatcellpm{3.0}{1.0} & \heatcellpm{8.0}{2.4} & \heatcellpm{7.0}{1.7} & \heatcellpm{2.5}{0.9} \\
\end{tabular}%
}
\end{table}

\subsection{Effect of Program Length on Training Configurations}
\label{app:ablation-length}

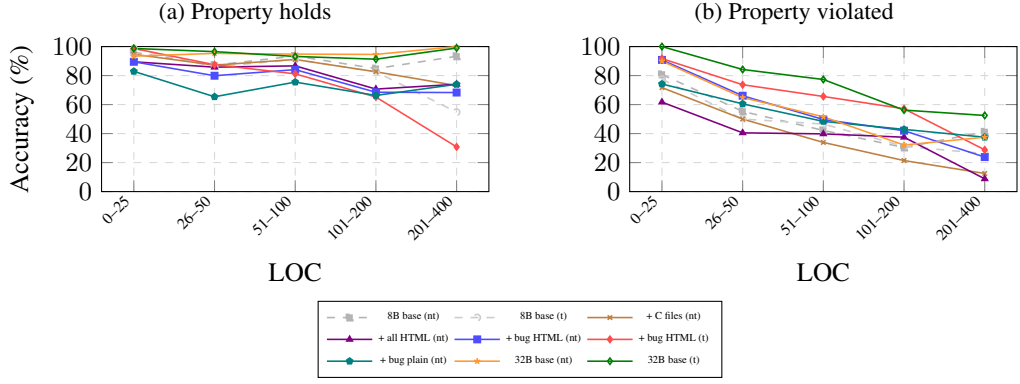
\begin{figure}[H]
\centering
\begin{subfigure}{0.48\columnwidth}
\centering
\subcaption{Property holds}
\begin{tikzpicture}
\begin{axis}[
    width=\textwidth,
    height=3.5cm,
    ylabel={Accuracy (\%)},
    xlabel={LOC},
    ymin=0, ymax=100,
    xtick={1,2,3,4,5},
    xticklabels={0--25,26--50,51--100,101--200,201--400},
    xticklabel style={font=\tiny, rotate=45, anchor=east},
    grid=major,
    grid style={dashed, gray!30},
    every axis plot/.append style={semithick, mark size=1.2pt},
    legend to name=ablationlegend,
    legend style={font=\fontsize{4}{5}\selectfont, legend columns=3},
]
\addplot[color=gray!60, mark=square*, dashed] coordinates {(1,96.1) (2,87.2) (3,93.5) (4,84.8) (5,93.3)};
\addplot[color=gray!40, mark=o, dashed] coordinates {(1,98.7) (2,86.0) (3,91.1) (4,82.6) (5,54.8)};
\addplot[color=brown, mark=x] coordinates {(1,94.7) (2,87.2) (3,91.1) (4,82.6) (5,73.1)};
\addplot[color=violet, mark=triangle*] coordinates {(1,89.5) (2,85.8) (3,86.7) (4,70.7) (5,74.0)};
\addplot[color=blue!70, mark=square*] coordinates {(1,89.5) (2,79.9) (3,84.1) (4,68.5) (5,68.3)};
\addplot[color=red!70, mark=diamond*] coordinates {(1,98.7) (2,87.5) (3,81.2) (4,65.2) (5,30.8)};
\addplot[color=teal, mark=pentagon*] coordinates {(1,82.9) (2,65.4) (3,75.5) (4,66.3) (5,74.0)};
\addplot[color=orange!80, mark=star] coordinates {(1,93.4) (2,95.3) (3,94.8) (4,94.6) (5,100.0)};
\addplot[color=green!50!black, mark=diamond] coordinates {(1,98.7) (2,96.5) (3,93.2) (4,91.3) (5,99.0)};
\legend{8B base (nt), 8B base (t), + C files (nt), + all HTML (nt), + bug HTML (nt), + bug HTML (t), + bug plain (nt), 32B base (nt), 32B base (t)}
\end{axis}
\end{tikzpicture}
\end{subfigure}
\hfill
\begin{subfigure}{0.48\columnwidth}
\centering
\subcaption{Property violated}
\begin{tikzpicture}
\begin{axis}[
    width=\textwidth,
    height=3.5cm,
    xlabel={LOC},
    ymin=0, ymax=100,
    xtick={1,2,3,4,5},
    xticklabels={0--25,26--50,51--100,101--200,201--400},
    xticklabel style={font=\tiny, rotate=45, anchor=east},
    grid=major,
    grid style={dashed, gray!30},
    every axis plot/.append style={semithick, mark size=1.2pt},
]
\addplot[color=gray!60, mark=square*, dashed] coordinates {(1,80.8) (2,55.3) (3,42.2) (4,30.4) (5,41.2)};
\addplot[color=gray!40, mark=o, dashed] coordinates {(1,77.5) (2,50.0) (3,46.6) (4,31.2) (5,26.2)};
\addplot[color=brown, mark=x] coordinates {(1,71.7) (2,50.0) (3,33.9) (4,21.4) (5,12.5)};
\addplot[color=violet, mark=triangle*] coordinates {(1,61.7) (2,40.5) (3,39.8) (4,37.5) (5,8.8)};
\addplot[color=blue!70, mark=square*] coordinates {(1,90.8) (2,66.1) (3,49.7) (4,42.0) (5,23.8)};
\addplot[color=red!70, mark=diamond*] coordinates {(1,91.7) (2,73.7) (3,65.6) (4,57.1) (5,28.8)};
\addplot[color=teal, mark=pentagon*] coordinates {(1,74.2) (2,60.5) (3,48.4) (4,42.9) (5,37.5)};
\addplot[color=orange!80, mark=star] coordinates {(1,90.0) (2,64.8) (3,51.6) (4,32.1) (5,37.5)};
\addplot[color=green!50!black, mark=diamond] coordinates {(1,100.0) (2,84.2) (3,77.3) (4,56.2) (5,52.5)};
\end{axis}
\end{tikzpicture}
\end{subfigure}

\vspace{0.1cm}
\centering\ref{ablationlegend}

\caption{Accuracy on our benchmark by program length for selected training configurations. (a) Holds accuracy degrades modestly with training on bug traces; the sharp drops at 201--400 LOC for thinking-enabled models reflect output budget exhaustion on longer programs (the model's chain-of-thought reasoning times out before producing a verdict). (b) Bug trace CPT with thinking (red) improves violation detection across all length bins, but the length-decay pattern persists: accuracy drops from 91.7\% (0--25 LOC) to 57.1\% (101--200 LOC), compared to the base model's drop from 80.8\% to 30.4\%. ``nt'' = no thinking, ``t'' = thinking.}
\label{fig:ablation-length}
\end{figure}

Figure~\ref{fig:ablation-length} shows accuracy by program length for selected training configurations, split by holds and violated. As with the baseline (Figure~\ref{fig:length-decay-main}), holds accuracy remains relatively stable across length bins. On the violation side, bug trace CPT with thinking lifts the curve uniformly, with the largest absolute gains in the 101--200 LOC bin where the baseline is weakest.

\subsection{Soteria Trace Example}
\label{app:soteria-trace}

The following shows a real example from our training data: Soteria's output for an uninitialized variable read, in both the original HTML format (used by \texttt{cpt-html-*} configurations) and the plain-text format (used by \texttt{cpt-bugs-*} and \texttt{soteria-*-plaintext} configurations). The plain-text conversion removes HTML tags, CSS classes, log-level prefixes, and decodes HTML entities while preserving the semantic content.

\textbf{Program (from CodeParrot):}
\begin{lstlisting}[language=C, basicstyle=\small\ttfamily, breaklines=true]
void exit(int status);

int main() {
  int x;

  if(x==10)
    exit(1);

  assert(x!=10);
}
\end{lstlisting}

\textbf{HTML trace (excerpt):}

\begin{lstlisting}[language={}, basicstyle=\fontsize{5.5}{6.5}\selectfont\ttfamily, breaklines=true]
<div class="log-msg DEBUG">Executing statement: {
  signed int x;
  if (rvalue(x) == 10)
    function_decay(exit)(1);
  else
    ;
  assert(rvalue(x) != 10);
}</div>
<div class="log-msg DEBUG">STORE:
  ( (x_563, { kind = Uninit; ty = signed int }))</div>
<div class="log-msg DEBUG">Executing statement: if (rvalue(x) == 10)
  function_decay(exit)(1);
else
  ;</div>
<div class="log-msg TRACE">Using state to error!</div>
<div class="log-msg TRACE">Obtained summary: After_exec
  ({ args = []; pre = []; pc = [];
     post = { heap = []; globs = [] };
     ret =
     (Error (Accessing uninitialized memory,
             [Triggering operation: file_0537382.c:6:6-7]))
     })</div>
<div class="log-msg DEBUG">Bug is manifest!!</div>
\end{lstlisting}

\textbf{Plain-text trace (excerpt):}

\begin{lstlisting}[language={}, basicstyle=\footnotesize\ttfamily, breaklines=true]
Executing statement: {
  signed int x;
  if (rvalue(x) == 10)
    function_decay(exit)(1);
  else
    ;
  assert(rvalue(x) != 10);
}
STORE:
  ( (x_563, { kind = Uninit; ty = signed int }))
Executing statement: if (rvalue(x) == 10)
  function_decay(exit)(1);
else
  ;
Using state to error!
Obtained summary: After_exec
  ({ args = []; pre = []; pc = [];
     post = { heap = []; globs = [] };
     ret =
     (Error (Accessing uninitialized memory,
             [Triggering operation: file_0537382.c:6:6-7]))
     })
Bug is manifest!!
\end{lstlisting}

Soteria tracks variable \texttt{x} as \texttt{Uninit} in the store. When the program reads \texttt{x} in the condition \texttt{x==10}, Soteria detects an access to uninitialized memory and reports the error with the precise source location. ``Bug is manifest'' indicates the error occurs on all feasible execution paths.

\subsection{Informalized Trace Example}
\label{app:informalized-trace}

The following shows a real example from our informalized training data, generated by prompting a frontier LLM to analyze C code in natural language based on Soteria's symbolic execution output. Each training example contains the C code followed by a step-by-step analysis.

\textbf{Program:}
\begin{lstlisting}[language=C, basicstyle=\small\ttfamily, breaklines=true]
extern void __VERIFIER_error() __attribute__ ((__noreturn__));
void __VERIFIER_assert(int cond) {
  if (!(cond)) {
    ERROR: __VERIFIER_error();
  }
  return;
}

int main(void) {
  unsigned int x = 0;
  unsigned short N;

  while (x < N) {
    x += 2;
  }

  __VERIFIER_assert(x % 2);
}
\end{lstlisting}

\textbf{Informalized analysis (excerpt):}

\begin{lstlisting}[language={}, basicstyle=\fontsize{6.5}{7.5}\selectfont\ttfamily, breaklines=true]
Looking at this code, I need to trace through what happens in `main`:

1. `x` is declared as `unsigned int` and initialized to `0`.
2. `N` is declared as `unsigned short` but **never initialized** -- it holds an indeterminate value.

Then we hit the loop:
  while (x < N) { x += 2; }

The loop condition compares `x` (which is `0`) against `N`, but `N` has never been assigned any value. Reading `N` here is accessing uninitialized memory, which is undefined behavior in C.

Now let's suppose we get past the UB and consider the assertion:
  __VERIFIER_assert(x % 2);

Here's the thing -- `x` starts at `0` and is only ever incremented by `2`. So `x` remains even through every iteration of the loop. That means `x % 2` is always `0`, regardless of what `N` is or how many times the loop runs.

So `__VERIFIER_assert(0)` is passed in. Since `cond` is `0`, `!cond` is true, and we branch into the error label, calling `__VERIFIER_error()`.

So there are really two issues here:
1. **Uninitialized read of `N`** -- undefined behavior at the loop condition.
2. **Assertion violation** -- even ignoring the UB, `x` is always even, so `x % 2` is always `0`, and the assertion will always fail.

The reachability of the error is guaranteed regardless of the value of `N`.
\end{lstlisting}

The informalized trace presents the analysis in natural language with explicit reasoning about execution paths, variable states, and the root cause of the violation. Unlike the raw Soteria output, it uses no tool-specific notation (no \texttt{STORE}, \texttt{HEAP}, or block labels), making it more accessible but less precise.

\subsection{Qualitative Example: CPT Model vs.\ Base Model}
\label{app:cpt-qualitative}

The following shows a case from our benchmark where the best-performing 8B CPT model with thinking (\texttt{qwen3-8b-cpt-html-bugs-thinking}) correctly detects a memory safety violation that the base model with thinking misses. The program searches for the first zero element in a nondeterministically initialized array:

\begin{lstlisting}[language=C, basicstyle=\small\ttfamily, breaklines=true]
#define N 1024
int main() {
  int A[N], i;
  for (i = 0; i < N; i++) A[i] = __VERIFIER_nondet_int();
  for (i = 0; A[i] != 0; i++) { if (i >= N) break; }
  __VERIFIER_assert(i <= N);
}
\end{lstlisting}

\textbf{Base model with thinking (incorrect --- says OKAY, excerpt):}

\begin{lstlisting}[language={}, basicstyle=\fontsize{6.5}{7.5}\selectfont\ttfamily, breaklines=true]
<think>
The array A is declared as int A[N], which is a static array.
The code accesses A[i] in the loops. Since the array is allocated
on the stack, the pointers here are valid. The code doesn't have
any pointers that are dereferenced after being freed or invalid.
So no invalid dereferences here.
</think>
Final answer: OKAY
\end{lstlisting}

The base model incorrectly concludes that all array accesses are valid because the array is stack-allocated, without considering the actual index values.

\textbf{CPT model with thinking (correct --- says ERROR, excerpt):}

\begin{lstlisting}[language={}, basicstyle=\fontsize{6.5}{7.5}\selectfont\ttfamily, breaklines=true]
<think>
...
The loop body checks if i >= N. If i is less than N, it does
nothing. But if i reaches N, then the loop breaks. Wait, but
the loop condition is A[i] != 0. So when i is N, the loop
condition would check A[N], which is out of bounds. Wait, the
array A has indices from 0 to N-1. So if i reaches N, then
A[i] is accessing A[N], which is beyond the array's bounds.
That's a problem.

So during each iteration, the loop checks A[i] != 0. But if
i is N-1, then after incrementing, i becomes N. Then the loop
condition checks A[N], which is out of bounds. So this would
cause an out-of-bounds access.
...
</think>
Final answer: ERROR
\end{lstlisting}

The CPT model correctly identifies that when \texttt{i} reaches \texttt{N} after incrementing, the loop condition \texttt{A[i] != 0} accesses \texttt{A[N]}, which is beyond the array's bounds (indices 0 to N$-$1). This result is consistent across all runs: the base model answers incorrectly in all 8 evaluations (4 with thinking, 4 without), while the CPT model answers correctly in all 4. This reasoning about boundary conditions and array indexing mirrors how Soteria's symbolic execution systematically tracks array bounds and explores all feasible paths, including the worst case where every nondeterministic value is non-zero.




\end{document}